\documentclass[pre,twocolumn,nofootinbib,twoside,showpacs,superscriptaddress,tightenlines,floatfix]{revtex4-2}
\usepackage{dcolumn}
\usepackage{lipsum}
\usepackage{graphicx,amssymb,amsmath,epsf,bm}
\usepackage[utf8x]{inputenc}
\usepackage[T1]{fontenc}
\usepackage[normalem]{ulem}
\usepackage{color}
\usepackage{mathtools}
\usepackage{dsfont}
\usepackage{booktabs}
\usepackage{nicefrac}
\usepackage{hyperref}
\hypersetup{
    allcolors=blue,
    colorlinks=true,
    bookmarks=true,
    pdfpagemode=FullScreen,
}
\definecolor{nred}{RGB}{224,0,0}
\definecolor{nblue}{RGB}{28,130,185}
\definecolor{dgreen}{RGB}{38,238,21}
\definecolor{norange}{RGB}{230,120,20}

\newcommand{\Tr}{{\rm Tr}}

\begin{document}


\title{Conformal invariance constraints in the $O(N)$ models: a first study within the nonperturbative renormalization group}

\author{Santiago Cabrera}
\affiliation{Instituto de F\'isica, Facultad de Ciencias, Universidad de la
	Rep\'ublica, Igu\'a 4225, 11400, Montevideo, Uruguay
}%

\author{Gonzalo De Polsi}
\affiliation{Instituto de F\'isica, Facultad de Ciencias, Universidad de la
	Rep\'ublica, Igu\'a 4225, 11400, Montevideo, Uruguay
}%

\author{Nicol\'as Wschebor}
\affiliation{Instituto de F\'isica, Facultad de Ingenier\'ia, Universidad de la
Rep\'ublica, J.H.y Reissig 565, 11000 Montevideo, Uruguay
}%

\begin{abstract}
  The behavior of many critical phenomena at large distances is expected to be invariant under the full conformal group, rather than only isometries and scale transformations.
  When studying critical phenomena, approximations are often required, and the framework of the nonperturbative, or functional renormalization group is no exception. The derivative expansion is one of the most popular approximation schemes within this framework, due to its great performance on multiple systems, as evidenced in the last decades. Nevertheless, it has the downside of breaking conformal symmetry at a finite order. This breaking is not observed at the leading order of the expansion, denoted LPA approximation, and only appears once one considers, at least, the next-to-leading order of the derivative expansion ($\mathcal{O}(\partial^2)$) when including composite operators. In this work, we study the constraints arising from conformal symmetry for the $O(N)$ models using the derivative expansion at order $\mathcal{O}(\partial^2)$. We explore various values of $N$ and minimize the breaking of conformal symmetry to fix the non-physical parameters of the approximation procedure. We compare our prediction for the critical exponents with those coming from a more usual procedure, known as the principle of minimal sensitivity.
\end{abstract}

\maketitle

\section{Introduction}

Critical phenomena are hard to study theoretically. This is due to the fact that strong correlations among the degrees of freedom of the system emerge in this regime. This implies that an approach based on sub-dividing the system into several independent parts is problematic. Moreover, in general, perturbative techniques (around a Gaussian or a mean-field approximation) present problems related to the appearance of strong couplings and require, at the very least, technical sophistication to succeed. In this context, Wilson's Renormalization Group (RG) was specifically developed in the 70s \cite{Wilson:1971dc,Wilson:1973jj} as a framework for tackling systems undergoing strong correlations, such as those which exhibit critical phenomena. A modern version of this framework, more suitable for the implementation of approximation schemes beyond perturbation theory, was developed in the 90s, known as the Functional or Non-Perturbative Renormalization Group (FRG) \cite{Wetterich:1992yh,Ellwanger:1993kk,Morris:1993qb,Delamotte:2007pf,Dupuis:2020fhh}. 

At criticality, thermodynamic properties show power law behaviors, characterized by exponents which are common to very different systems. This is a consequence of the exhibited scale invariance from a mesoscopic scale onwards. This also explained universality, as was shown by Wilson \cite{Wilson:1973jj}. In addition to systems presenting invariance under scale transformations, Migdal and Polyakov first conjectured that also full conformal invariance was to hold at many critical phenomena \cite{Polyakov:1970xd,Migdal:1971xh}.

Nowadays, it is known that under rather general conditions conformal symmetry does indeed take place for 2D critical phenomena \cite{DiFrancesco:1997nk,Belavin:1984vu}, which has allowed to fully classify conformal theories and compute exactly critical properties in that dimension. Although there is no general proof of the realization of this symmetry in the very relevant 3D case, this is expected to be the case for the vast majority of physical systems. In particular, it was shown to be the case for a few models such as the Ising or $\mathbb{Z}_2$ model \cite{delamotte2016scale} and the $O(2)$, $O(3)$ and $O(4)$ models \cite{DePolsi2019} (and it is expected to be valid also for all $O(N)$ models) once isometries, scale invariance and a few other mild assumptions are required. This has a relevance which is not just purely academic as $O(N)$ models are in the same universality classes as many physically relevant systems such as pure substances and uni-axial antiferromagnets ($N=1$), planar magnets and Helium-4 fluid-superfluid $\lambda$-transition  ($N=2$), isotropic magnets ($N=3$), among many other examples.

Considerable effort has been dedicated to the study of critical phenomena using very different techniques. These include Monte-Carlo simulations \cite{Hasenbusch:2019jkj,Hasenbusch2005,Hasenbusch:2021tei}, as well as other standard methods such as the $\epsilon$-expansion, which are still being pushed to higher and higher orders \cite{Schnetz:2016fhy,Kompaniets:2017yct,Shalaby:2020faz,Abhignan06,Shalaby:2020xvv}. There are also some other more recent ones such as the Conformal Bootstrap which, assuming the realization of conformal symmetry at criticality, has been able to reach a remarkable precision in the computation of some universal quantities in the Ising model universality class, and have also produced very good results for some $O(N)$ models \cite{El-Showk:2014dwa,ElShowk:2012ht,Kos:2014bka,chester2019carving,meneses2019structural}. In the context of the FRG, critical phenomena have been studied mainly (but not exclusively) by means of an approximation technique known as the derivative expansion (DE, see below for some detail), which yielded very precise and accurate results. In some cases, it provides the most precise computations in the reported literature at the moment \cite{DePolsi2020,DePolsi2021,Peli:2020yiz,Sanchez2023,Rennecke2024,Sanchez2024}.

The DE of the FRG is a controlled approximation technique which consists in considering a general \textit{ansatz} for the effective action (Gibbs free energy) including terms with a finite number of derivatives of the field, which is equivalent to expanding vertices around zero momentum. When one is interested in thermodynamic properties which are obtained from the zero momentum regime, this approximation is a very powerful tool. Moreover the FRG procedure regulates the infrared physics, allowing the expansion around zero momenta that turns out to be a controlled approximation with a small parameter of order $\sim$1/4 \cite{Balog2019}. This allows for the introduction of error bars \cite{DePolsi2020,DePolsi2021,Chlebicki:2022pxm,Sanchez-Villalobos:2024vmd}. It is worth emphasizing that, in order, for the good properties of convergence to manifest and for the approximation to be controlled, it is necessary to fix properly the spurious dependence on the regularization process which appears due to the fact that approximations are being implemented \cite{DePolsi2022}. An appropriate and pragmatical criterion, known as the Principle of Minimal Sensitivity (PMS), has been exploited to fix this dependence. It is based on the fact that, since physical quantities should not be dependent on auxiliary parameters, the best results are to be expected when the dependence of observables on these auxiliary parameters is as small as possible.

Despite the great success of the DE in computing critical properties, it was not until recently that conformal symmetry has been exploited as a source of information in this context. As it turns out, the DE approximation breaks conformal invariance at a given finite order. In recent works, this fact was used to attempt to fix the spurious dependence on the regularization procedure \cite{Balog2020,Delamotte2024}, by requiring conformal invariance to be fulfilled as much as possible. This led to another criterion for fixing these parameters called Principle of Maximal Conformality (PMC). The main outcome of these studies is that using either PMS or PMC for fixing the spurious dependence on nonphysical parameters yields 
compatible results (i.e. within error bars). However, all the studies done so far concerning the role of conformal invariance in the FRG context  analyze only the Ising model universality class.

As it was stated, these previous studies set the ground for a more physical justification of the PMS, as its seems to coincide with the best realization of conformal symmetry. The goals of this paper is two-fold. The first purpose is to extend the previous work reported in \cite{Delamotte2024} to the $O(N)$ universality classes. The aim is to further test the conjectured equivalence of the PMS and PMC. This requires considering correlation functions including composite operators that respect the $O(N)$ symmetry. These correlation functions enclose the information of several perturbations around the fixed point that include, among others, the most relevant one associated with the critical exponent $\nu$ and the first correction associated to the critical exponent $\omega$.

Our second goal is to compare two slightly different implementations of the DE, that we denote as the \textit{full ansatz} and \textit{strict version} to test their equivalence within error bars. These two implementations will be explained in more detail below, but in a nutshell, they differ in the fact that the strict version drops all higher order terms while the full version retains some of them.\footnote{The full version of the DE preserves some kinematic symmetry properties of the various vertices which are not present in the strict implementation.} Given the fact that there is no rigorous proof of the convergence of the DE, this is a consistency check of its good behavior.

The outline of this article is the following. We start with a brief introduction to the FRG and its essential features in the context of $O(N)$ models and in the presence of sources for both the field and composite operators. In particular, we describe the DE approximation in this framework. We then discuss symmetries in this framework focusing on dilatation and special conformal transformations. Next we consider the DE at next-to-leading order to study the $O(N)$ models for different values of $N$. Since we need to employ an approximation scheme, which breaks conformal symmetry, we find that, in the presence of a source for composite operators, the Ward identity for special conformal transformations is violated. The minimization of this breaking is then used as a way of fixing the spurious dependence on nonphysical parameters, giving rise to the PMC criterion. Next we compare the results coming from both the PMS and PMC criteria for the critical exponents $\nu$ and $\omega$. Finally, we draw our conclusions.

\section{Functional renormalization group essentials}\label{secFRGON}

\subsection{Some elements of functional renormalization group}
The FRG underlying idea is to decouple long-wavelength fluctuations, with respect to an artificially introduced running scale $k$. This allows for the short-wavelength fluctuations to be incorporated gradually into the system's description in a controlled manner. One possible way to implement this idea is to modify the action of the theory by adding a term quadratic in the fields and dependent on an auxiliary scale $k$ \cite{Polchinski:1983gv} that effectively freezes long-wavelengths fluctuations while leaving the short-wavelength (or fast, in momentum space) modes unchanged:
\begin{equation}\label{deltaSDirect}
	\begin{split}
		\Delta S_k[\vec\varphi]&=\frac 1 2 \int_{x,y}\frac{\varphi_a(x) R_k(|x-y|^2)\varphi_a(y)}{2}\\
		&=\frac 1 2 \int_{q}\frac{\varphi_a(q) R_k(q^2)\varphi_a(-q)}{2},
	\end{split}
\end{equation}
where the same notation is used for functions and their Fourier-transformed versions. We have also introduced the notations $\int_x\equiv\int d^dx$ and $\int_q\equiv\int \frac{d^dq}{(2\pi)^d}$ for simplicity. In equation \eqref{deltaSDirect} and henceforth, repeated indices are to be considered as being summed over unless otherwise stated. 
To preserve translation, rotational and $O(N)$ invariance along the flow $R_k(q^2)$ has to present the following general profile
\begin{equation}\label{eq:RegProf}
R_k(q^2)=\alpha Z_k k^2 r(q^2/k^2),
\end{equation}
with the function $r$ decaying faster than any power law for large momentum and tending to one for vanishing momentum $q$.  In equation \eqref{eq:RegProf}, $Z_k$ is a field renormalization factor and $\alpha$ plays the role of an auxiliary parameter which sets the scale of the regulator and is used to optimize computations. 

With the addition of this regulator term to the action, one arrives at a scale dependent functional integral for the partition function of the form:
\begin{equation}\label{functIntegral}
	e^{W_k[J,K]}=\int \mathcal{D}\varphi e^{-S[\vec{\varphi}]-\Delta S_k[\vec{\varphi}]+\int_x J_a(x)\varphi_a(x)+\int_x K(x)\mathcal{O}(x)},
\end{equation}
where we consider for the purpose of this work a source for the fields, $J_a(x)$, and a source $K(x)$ for generic $O(N)$ invariant local operators, that we denote here as $\mathcal{O}(x)$, following \cite{Ellwanger:1993kk,Berges:2000ew,Pawlowski:2005xe,Dupuis:2020fhh,Rose:2015bma,Rose:2021zdk}. 

Starting from the function integral \eqref{functIntegral}, a Legendre transformation of the Helmholtz free energy $W_k[\vec{J},K]$ with respect of $J_a$ defines a scale-dependent effective action $\Gamma_k[\phi,K]$ given by:
\begin{equation}
\label{eq_legendre}
	\Gamma_k[\vec{\phi},K]+\Delta S_k[\vec{\phi}]=-W_k[\vec{J},K]+\int_x J_a(x)\phi_a(x).
\end{equation}
Notice that in defining the scale dependent effective action $\Gamma_k[\vec{\phi},K]$, we have extracted the regulator term. This is done in order for $\Gamma_k[\phi,K]$ to properly interpolate between the microscopic action defined at an ultraviolet scale $k=\Lambda$ and the actual effective action, or Gibbs free energy, at $k=0$. We recall that, therefore, the order parameter $\phi_a(x)$ is defined as:
\begin{equation}\label{orderParam}
\frac{\delta W_k[\vec{J},K]}{\delta J_a(x)}=\langle\varphi_a(x)\rangle_{\vec{J},K}\equiv\phi_a(x).
\end{equation}

Following a standard procedure \cite{Delamotte:2007pf}, a flow equation for the FRG effective action with respect to the scale $k$ can be derived and it is given by:
\begin{equation}\label{WettEq}
	\partial_t \Gamma_k[\vec{\phi},K]=\frac{1}{2}\int_{x,y}\partial_t R_k(x,y)G_{k,aa}[x,y],
\end{equation}
where $t\equiv \log (k/\Lambda)$, $\Lambda$ and the full propagator $G_{k,ab}[x,y]$ is as usual:
\begin{equation}
\left\langle\varphi_a(x)\varphi_b(y)\right\rangle_c=\frac{\delta^2 W_k[\vec{J},K]}{\delta J_a(x)\delta J_b(y)}=\Big(\frac{\delta^2\Gamma_k[\vec{\phi},K]}{\delta\phi_a\delta\phi_b}+\delta_{ab}R_k\Big)_{x,y}^{-1}.
\end{equation}
In order to study the critical properties of the system and to look for a fixed point of the RG, it is important to introduce dimensionless variables which amounts to measuring all quantities in unites of the regulator scale $k$. We thus introduce the following variables:
\begin{equation}
\label{dimensionless}
   \begin{split}
       \tilde{x}&=k x,\\
       \tilde{\phi}_a(\tilde{x})&=Z_k^{\nicefrac{1}{2}}k^ {\frac{2-d}{2}}\phi_a(x),\\
       \tilde{K}(\tilde{x})&=Z_k^{(\mathcal{O})}k^{-d}K(x),
   \end{split} 
\end{equation}
where we denote with a tilde dimensionless variables. The renormalization factors $Z_k$ and $Z_k^{(\mathcal{O})}$ must be fixed with appropriate renormalization conditions to be made explicit below.

Implementing this change of variables leads to the dimensionless version of Eq.~\eqref{WettEq}:
\begin{equation}
\begin{aligned}
    \label{WettEqDless}
\partial_t \Gamma_k[\tilde{\phi}_a,\tilde{K}]=&\int_{\tilde{x}}\frac{\delta \Gamma_k}{\delta \tilde{\phi}_a(\tilde{x})}\big(\tilde{x}^\nu\tilde{\partial}_\nu+D_\varphi\big)\tilde\phi_a(\tilde{x})
\\
&-\int_{\tilde{x}}\frac{\delta \Gamma_k}{\delta \tilde{K}(\tilde{x})}\big(\tilde{x}^\nu\tilde{\partial}_\nu+D_{\mathcal{O}}\big)\tilde{K}(\tilde{x})
\\
+\frac{1}{2}\alpha\int_{\tilde{x},\tilde{y}}&\bigg[(d+2-\eta_k)r(|\tilde{x}-\tilde{y}|)
\\
&+|\tilde{x}-\tilde{y}|r'(|\tilde{x}-\tilde{y}|)\bigg]\tilde{G}_{k,aa}[\tilde{x},\tilde{y}].
\end{aligned}
\end{equation}
We have introduced the running anomalous dimension $\eta_k$ defined as $\partial_t Z_k=-\eta_k Z_k$. Similarly, we introduced the running dimensions of the field $\phi$, $D_\varphi=(d-2+\eta_k)/2$ and of the operator $\mathcal{O}$, 
$D_\mathcal{O}=\partial_t \log(Z_k^{(\mathcal{O})})$. Notice that, typically, $\eta_k$, $D_\varphi$ and $D_\mathcal{O}$ depend on $k$. However, at a fixed point of the RG, $\eta_k^*=\eta$ becomes the 
anomalous dimension of the field, and $D_\varphi$ and $D_\mathcal{O}$ become, respectively, the scaling dimension of the field $\phi$ and the operator $\mathcal{O}$.

Up to now, the presentation has been general and describes the full flow of the effective action. However, in the context of this work, we will be only interested in the linearized flow of the effective action around the fixed point for vanishing source $\tilde{K}$ given by ${\partial_t \Gamma_k[\tilde{\phi}_a,\tilde{K}=0]=0}$. As will be explained below, this linearized flow can be obtained from the fixed point equation but including the source $K$ at linear order. As a consequence we will only take into consideration the linear dependence of the effective action with $K$. 
As shown in Ref.~\cite{Delamotte2024}, FRG flow equations in the presence of sources for composite operators have a triangular property. The flow of the effective action at zero sources does not depend on the effective action at non-zero sources. In a similar way, the flow of the effective action expanded at linear order in the source $K$ does not depend on terms corresponding to higher powers of the source. As a consequence, one can limit the effective action to linear order in $K$ and this does not imply an additional approximation.
Of course, a more general study could be performed keeping higher powers of $K$ but this goes beyond the scope of the present work.

\subsection{The derivative expansion}

Except for very few examples, such as the Large $N$ limit of the $O(N)$ models, approximation techniques are required to analyze an equation such as Eq. \eqref{WettEq}. One of the most frequently used approximations in the FRG context is the DE that consists in considering an \textit{ansatz} for the scale-dependent effective action with all the possible terms including up to a given number of derivatives and to consistently project onto the space of solutions compatible with the considered \textit{ansatz}. A given order of the approximation with up to $s$ derivatives is referred to as the order $\mathcal{O}(\partial^s)$ of the DE.

It is now well established that the first orders of this approximation scheme already yield very precise results. Several examples of this fact can be found in the literature \cite{Berges2002,Canet2003b,Canet2003,Balog2019,DePolsi2020,DePolsi2021}, but we refer the reader to a recent review of the FRG formalism \cite{Dupuis:2020fhh} for further information.

To clarify the preceding ideas we present the example of the first order of this procedure $\mathcal{O}(\partial^0)$, also known as the local potential approximation (LPA) and the following order $\mathcal{O}(\partial^2)$, which is the one to be considered in this article. We will consider both approximations in the presence of a source for composite operators. The LPA consists of the following ansatz:
\begin{equation}\label{eq:ch2:isingLPA}
\Gamma_k[\phi_a,K]=\int_x \biggl\{\frac{1}{2}\big(\nabla\phi_a\big)^2+U_{0}(\rho)+K(x)U_{1}(\rho)\biggr\},
\end{equation}
where we introduced the $O(N)$ invariant $\rho\equiv\phi_a\phi_a/2$ and an unrenormalized kinetic term is taken into consideration to account for interactions. Notice that, as previously stated, we have only kept the linear dependence on the source $K(x)$. In principle, the source $K$ could be coupled to very general local operators. For simplicity, in Eq.~(\ref{eq:ch2:isingLPA}) and below we consider only operators that are scalars under isometries and that are $O(N)-$invariant. In the LPA case (that is, excluding derivative terms), this reduces the ansatz to a function $U_{1}(\rho)$ of the single $O(N)-$invariant scalar without derivatives, $\rho$. It is useful to mention nevertheless that there is another very important operator without derivatives that can be treated in an exact way but is not $O(N)-$invariant: the operator $\mathcal{O}=\phi_a(x)$. In that case, all $\Gamma^{(n,1)}$ are zero except for $n=1$. As a consequence, all loop terms are zero and one concludes only the expected and trivial result $D_{\mathcal{O}}=D_{\varphi}$. That is, as expected, the introduction of this linear operator does not give further information with respect to treating only, as usual, the $\Gamma^{(n,0)}$ vertices.

The next-to-leading order $\mathcal{O}(\partial^2)$ \textit{ansatz} takes, then, the following form:
\begin{equation}\label{ansatz}
\begin{split}
\Gamma_k[\phi,K]&=\int_x\Big\{U_0(\rho)+\frac{Z_0(\rho)}{2} (\nabla\phi_a)^2+\frac{Y_0(\rho)}{4}(\nabla\rho)^2\\+K(x&) \bigg(U_1(\phi)+\frac{Z_1(\rho)}{2}(\nabla\phi_a)^2+\frac{Y_1(\rho)}{4}(\nabla\rho)^2)\bigg)\\
&- \Upsilon(\phi) \partial^2 K(x) \Big\}.
\end{split}
\end{equation}
Notice that we omit the $k$ dependence of the \textit{ansatz} functions $U_0$, $U_1$, etc. in order to ease the notation. Here, again, we assumed the local operator to be scalar under isometries and $O(N)-$invariant.

As was mentioned, the DE is known to produce results that are consistent with other theoretical methods and that seem to converge nicely as one considers higher orders. Partly due to the relatively recent nature of the FRG, it was not until long ago that an explanation for this was found, although a rigorous proof is still not available. Succinctly, the infrared behavior of the correlation functions is regularized by the presence of the regulator which ensures the smoothness of the vertices as a function of momenta, which, in turn, allows for this expansion around zero momentum to be controlled. Moreover, the FRG has a one-loop structure with the particular characteristic that the diagrams include the factor $\partial_t R_k(q)$ in the numerator which further suppresses the large momentum contributions in view of the properties that $R_k(q)$ satisfies. This restriction of the domain of momenta contributing to the integrals implies that it is possible to expand vertices at small momenta, both for internal and external momenta and that the flow of vertices with small momenta (with respect to the regulator scale $k$) are insensitive to the high momenta behaviors. 

Out of criticality, the radius of convergence of an expansion around small momenta is related to the nearest pole in the complex plane of $p^2$. Since the regulator is a mass-like term which takes the system out of criticality, this expansion is expected to be justified even if the original theory is at the critical regime. For the $O(N)$ models, the radius of convergence has been shown to be of the order $q^2/k^2\simeq 4$ \cite{Balog2019}. This is consistent with the numerical observations for the $O(N)$ models, at least up to order $\mathcal{O}(\partial^4)$ \cite{DePolsi2020,DePolsi2021,Peli:2020yiz}, which is the highest order of the DE implemented in the literature for this model to date. Likewise, for the Ising model universality class, the DE has been implemented up to order $\mathcal{O}(\partial^6)$ \cite{Balog2019} without changes in the observed behavior. These studies allowed to understand why this approximation scheme, in the context of the FRG and for $O(N)$ models, has rather good convergence properties with a small parameter of order $\sim \frac{1}{4}$ or smaller.

To extract the flow of the various functions, one must expand the flow equations for the various vertices in momenta. For example, to determine the flow of $Z_0$ at order $\mathcal{O}(\partial^2)$ one calculates the vertices with the \textit{ansatz} given in \eqref{ansatz}, with $K$ set to zero for simplicity. Then, one inserts these vertices into the evolution equation for $\Gamma_{k,ab}^{(2,0)}(p)$:
\begin{equation}\label{flowGamm2}
        \partial_t\Gamma_{k,n_1 n_2}^{(2,0)}(p)=\text{Tr}\Big[ \dot{R}_k \cdot G\cdot H^{(2)}_{\cdot n_1n_2\cdot}(p) G\Big],
\end{equation}
where we employed a matrix notation to avoid an even more cumbersome expression, and omitted writing the scale $k$ in the propagator for the same reason. In ~\eqref{flowGamm2} we also introduced the notation
\begin{equation}
    \begin{split}
        H^{(2)}_{an_1n_2b}(q,p,q')&=-\frac{1}{2}\Big[\Gamma^{(4,0)}_{an_1bn_2}(q,p,q')\\+\Gamma^{(3,0)}_{an_1c}&(q,p)G_{cd}(q+p)\Gamma^{(3,0)}_{dbn_2}(q+p,q')\\+\Gamma^{(3,0)}_{bn_1c}&(q',p)G_{cd}(q'+p)\Gamma^{(3,0)}_{dan_2}(q'+p,q)\Big],
    \end{split}
\end{equation}
where in general, the function $H^{(m)}_{an_1\dots n_mb}(q,p_1,\dots,p_{m-1},q')$ stands for the different diagrammatic contributions that arise when taking functional derivatives (and then performing a Fourier-transform) in the FRG equation \eqref{WettEq}. The function $G_{ab}(q)$ stands for the Fourier-transform of $G_{ab}(x,0)$ evaluated at a uniform field configuration and $\Gamma_{a_1\dots a_m}^{(m,0)}(q_1,\dots,q_{m-1})$ for the Fourier-transform of $\Gamma_{a_1\dots a_m}^{(m,0)}(x_1,\dots,x_{m-1},0)$ evaluated at a uniform field configuration where we denote
\begin{equation}
	\begin{split}
		\Gamma^{(n,m)}_{k,a_1\dots a_n}&(x_1,\cdots,x_n;y_1,\cdots,y_m)=\\& \frac{\delta^{n+m} \Gamma_k}{\delta\phi_{a_1}(x_1)\cdots\delta\phi_{a_n}(x_n)\delta K(y_1)\cdots\delta K(y_m)}.
	\end{split}
\end{equation}

The \textit{left hand side} of equation \eqref{flowGamm2} is:
\begin{equation}
	\begin{split}
            \partial_t\Gamma_{k,a_1 a_2}^{(2,0)}&(p)=\delta_{a_1 a_2}\Big(\partial_t U_0^{(1)}(\rho)+p^2\partial_tZ_0(\rho)\Big)\\+&\phi_{a_1}\phi_{a_2}\Big(\partial_t U_0^{(2)}(\rho)+\frac{p^2}{2}\partial_tY_0(\rho)\Big).
	\end{split}
\end{equation}
Hence, to extract the flow of $Z_0(\rho)$ one needs to isolate the contribution proportional to $\delta_{a_1 a_2}p^2$ from the \textit{right hand side} of equation \eqref{flowGamm2}.

At this point, it becomes clear that one can simply plug-in the \textit{ansatz} for the vertices, expand the propagator $G_{ab}(q+p)$ around $q$ and obtain an expression in terms of powers of $p^2$. However, when doing so straightforwardly, one includes terms that come from the product of vertices which are of order $q^4$ or $p^2q^2$. This type of contributions are in fact, of the same order as contributions coming from a $\partial^4$ term in the \textit{ansatz} which was already neglected and can be dropped accordingly in a consistent way \textit{a priori}. In the literature there are, at least, two different implementations of the DE technique. We will refer to the approximation where this kind of terms is neglected as the \textit{strict} implementation of the DE and the one where those terms are kept the \textit{full ansatz} one. It is worth mentioning that the full implementation preserves some kinematic symmetries (of the exact theory) which are lost when using the strict implementation.

\subsection{Eigenperturbations of flow equations around the fixed point}
\label{sec_eigenperturbations}

The relationship between scaling operators and the eigenperturbations of the RG flow at the fixed point was discussed previously in Ref. \cite{Delamotte2024}. Here we recall this relation in order to apply it to $O(N)-$invariant models. To this end, consider a fixed point solution of equation \eqref{WettEqDless} at zero source for the composite operators, say $\Gamma_*[\tilde\phi,\tilde K=0]$, and let us study the flow around it. One can define: 
\begin{equation}
    \gamma_k[\tilde{\phi}_a]=\Gamma_k\big[\tilde{\phi}_a,K=0\big]-\Gamma_*\big[\tilde{\phi}_a,K=0\big],
\end{equation}
substitute it into equation \eqref{WettEqDless} and evaluate its flow near the fixed point solution at linear order in $\gamma_k[\tilde{\phi}_a]$. To simplify notation, we will drop the tilde on quantities since we will be only interested in dimensionless variables from this point onwards. This procedure yields,\footnote{It is worth emphasizing that in equation \eqref{perturbaroundfixed}, the term $\dot{R}_k$ stands for the  dimensionless version of it, made explicit in equation \eqref{WettEqDless}.} 

\begin{equation}\label{perturbaroundfixed}
 \begin{split}
\partial_t \gamma_k[\vec{\phi}]+\int_x&\gamma_{k,a}^{(1)}[\vec{\phi};x] \big(D_\varphi+x_\mu \partial_\mu\big) \phi_a(x)
\\&=\frac{1}{2}\Tr\big[ \dot{R}_k G_{k}\cdot\gamma_{k}^{(2)}[\vec{\phi}]\cdot G_{k}\big],
 \end{split}
\end{equation}
where, once again, the latin letters indicate the $O(N)$ color indexes the trace stands for a volume integral to simplify notation. In Eq.~\eqref{perturbaroundfixed} we introduced the scaling dimension of the field $\varphi$, $D_{\varphi}=(d-2+\eta^*)/2$. To avoid having to consider linear perturbations on the running anomalous dimension $\eta_k$, we adopt here a renormalization scheme where $\eta_k$ is kept at its fixed point value, $\eta_k=\eta^*, \forall k$. Close to the fixed point, this is a valid renormalization scheme, see \cite{Morris:1993qb,Delamotte2024}.

Considering equation \eqref{perturbaroundfixed} one can look for eigenperturbations, namely solutions of the form:
\begin{equation}\label{eigenpert}
	\gamma_k[\vec{\phi}]=\exp(\lambda t) \hat{\gamma}[\vec{\phi}],
\end{equation}
where $\hat{\gamma}[\vec{\phi}]$ is $t-$independent. This leads to the following eigenvalue equation:
\begin{equation}\label{perturbaroundfixed_bis}
	\lambda \hat{\gamma}[\vec{\phi}]=\frac{1}{2}\Tr\Big[ \dot{R}_k G\cdot \hat{\gamma}^{(2)}[\vec{\phi}]\cdot G\Big]-\int_x\hat{\gamma}_a^{(1)}[\vec{\phi};x] \big(D_\varphi+x_\mu \partial_\mu\big) \phi_a.
\end{equation}

Let us now relate the linearized flow around the fixed point of the RG to the flow in the presence of the source $K(x)$. Let us start from \eqref{WettEqDless} and consider the evolution of the linear perturbation around zero sources:
\begin{equation}\label{gammaHat}
	\hat{\Gamma}[\vec{\phi}]=\int_x\Gamma_k^{(0,1)}\Big|_{K=0}=-\Big\langle\int_x\mathcal{O}(x)\Big\rangle_{J,K=0}.
\end{equation}
It has the following flow equation: 
\begin{equation}\label{scalingEvol}
	\begin{split}
		\partial_t \hat{\Gamma}_k[\vec{\phi}]+&\big(D_\mathcal{O}-d\big)\hat{\Gamma}_k[\vec{\phi}]=-\int_x\hat{\Gamma}_a^{(1)}[\vec{\phi};x] \big(D_\varphi+x_\mu \partial_\mu\big) \phi_a\\&+\frac{1}{2}\Tr\big[ \dot{R}_k G_{k}\cdot\hat{\Gamma}_{k}^{(2)}[\vec{\phi}]\cdot G_{k}\big],
	\end{split}
\end{equation}
which, evaluated at the generalized fixed point including sources, where $\partial_t\hat{\Gamma}^*[\vec{\phi}]=0$, reduces to:
\begin{equation}\label{scalingEvolFP}
	\begin{split}
		\big(D_\mathcal{O}-d\big)\hat{\Gamma}^*[\vec{\phi}]&=-\int_x\hat{\Gamma}_a^{*(1)}[\vec{\phi};x] \big(D_\varphi+x_\mu \partial_\mu\big) \phi_a\\&+\frac{1}{2}\Tr\big[ \dot{R}_k G_{k}\cdot{\hat{\Gamma}_{k}}^{*(2)}[\vec{\phi}]\cdot G_{k}\big].
	\end{split}
\end{equation}
where, at the fixed point, $D_\mathcal{O}=D_\mathcal{O}^*$ is the scaling dimension of the operator $\mathcal{O}$. It becomes evident now, by comparing equations \eqref{perturbaroundfixed_bis} and \eqref{scalingEvolFP}, that at the fixed of the full effective action in the presence of sources point $\Big\langle\int_x\mathcal{O}(x)\Big\rangle_{J,K=0}$ is an eigenperturbation around the fixed point of the effective action {\it without} sources. Moreover, the stability matrix eigenvalue $\lambda$ given in equation \eqref{eigenpert} and the scaling dimension $D_\mathcal{O}$ of the operator $\mathcal{O}(x)$ are related by:
\begin{equation}
	\lambda=D_\mathcal{O}-d.
\end{equation}
In fact, any linear perturbations around the fixed point can be written as a linear combination of eigenperturbations. Accordingly, without loss of generality we will consider the source $K$ as being coupled to one such local eigenperturbation of the RG fixed point.

One should observe that equations \eqref{perturbaroundfixed} and \eqref{perturbaroundfixed_bis} can be generalized straightforwardly for any small perturbation around the fixed point, whether or not it is rotation or translation invariant, or even invariant under the internal symmetries of the fixed point.

\section{Conformal group symmetry within the functional renormalization group}\label{secSCSFRG}

In this section we discuss symmetries and their realization in the formalism of the FRG for the $O(N)$ model and in the presence of a source for both the field and local composite primary eigenoperators. 

\subsection{Ward identities for dilatation and conformal invariance}
\label{sec_WIDC}

We assume hereafter that the integrating measure in the functional integral is invariant under all the conformal transformations, namely translations, rotations,  dilatations and special conformal transformations. These last two transformations are infinitesimally realized by the following variations of the fields:
\begin{equation}\label{eqTransfField}
	\begin{split}
		\delta_{\rm dil}\,\varphi_a(x)&=\epsilon \big(x_\mu \partial_\mu+D_\varphi\big) \varphi_a(x),\\
		\delta_{\rm conf}\, \varphi_a(x)&=\epsilon_\mu \big(x^2\partial_\mu-2x_\mu x_\nu\partial_\nu-2x_\mu D_\varphi\big) \varphi_a(x).\\
	\end{split}
\end{equation}
In a similar manner, a \textit{primary} operator $\mathcal{O}(x)$ is a composite operator\footnote{As pointed out before, we only consider composite operators $\mathcal{O}(x)$ which are scalars under isometries and $O(N)-$invariant.} which transforms in the same way as the field but with a different scaling dimension $D_\mathcal{O}$:
\begin{equation}\label{eqTransfOperat}
	\begin{split}
		\delta_{\rm dil}\, \mathcal{O}(x)&=\epsilon \big(x_\mu \partial_\mu+D_{\mathcal{O}}\big) \mathcal{O}(x),\\
		\delta_{\rm conf}\, \mathcal{O}(x)&=\epsilon_\mu 
		\big(x^2\partial_\mu-2x_\mu x_\nu\partial_\nu-2x_\mu D_{\mathcal{O}}\big)\mathcal{O}(x).
	\end{split}
\end{equation}
This implies, in particular, that the operator $\mathcal{O}$ is associated to an eigenperturbation around the fixed point, as mentioned before, but the condition to be primary is more restrictive.

Starting from these transformation laws, inserting them into the functional integral \eqref{functIntegral} and performing the Legendre transformation described previously one arrives \cite{Rosten:2014oja,Sonoda:2015pva,delamotte2016scale,Rosten:2016zap,Balog2020} at the modified version of the Ward identities (in the presence of the regulator) for dilatations:
\begin{equation}\label{Warddilat}
	\begin{split}
		\int_x \Big\{\Gamma_{k,a}^{(1,0)}&(x) \big(D_\varphi+x_\mu \partial_\mu\big) \phi_a(x)\\-K&(x)\big(D_{\mathcal{O}}+x_\mu \partial_\mu\big) \Gamma_k^{(0,1)}(;x)\Big\}\\&=-\frac{1}{2}\int_{x,y}\partial_t R_k(x,y)G_{k,aa}[x,y],
	\end{split}
\end{equation}
and for special conformal transformations:
\begin{equation}\label{Wardconformal}
	\begin{split}
		\int_x\Big\{\Gamma_{k,a}^{(1,0)}&(x) \big(x^2\partial_\mu-2x_\mu x_\nu\partial_\nu-2x_\mu D_\varphi\big) \phi_a(x)\\-K(x)\big(&x^2\partial_\mu-2x_\mu x_\nu\partial_\nu-2x_\mu D_{\mathcal{O}}\big) \Gamma_k^{(0,1)}(;x)\Big\}\\=\frac{1}{2}&\int_{x,y}(x_\mu+y_\mu)\partial_t R_k(x,y)G_{k,aa}[x,y],
	\end{split}
\end{equation}
where it was used that
\begin{equation}
	(2d-2D_\varphi+2x^2\partial_{x^2})R_k (x^2)=\partial_t R_k (x).
\end{equation}

It is interesting to note that the modified dilatation Ward identity \eqref{Warddilat} is nothing but the fixed point equation obtained by setting $\partial_t \Gamma_k[\phi,K]=0$ in Eq. \eqref{WettEqDless}. Similarly, Eq. \eqref{scalingEvolFP} is related to the modified dilatation Ward identity of $\hat{\Gamma}[\phi]$, obtained by differentiating equation \eqref{Warddilat} with respect to $K(y)$, evaluating at $K=0$ and integrating over $y$.

Since we are only interested in the $K-$independent or linear dependence with $K$, we will limit ourselves to consider just the vertices $\Gamma^{(n,0)}|_{K=0}$ and $\Gamma^{(n,1)}|_{K=0}$.

One can consider the vertices $\Gamma^{(n,0)}|_{K=0}$ in momentum space. In order to do so we evaluate the vertices at uniform field and perform a Fourier-transform with respect to all but one of the coordinates, which is set as the origin by exploiting translation invariance. One proceeds similarly with $\Gamma^{(n,1)}|_{K=0}$. We employ in that case the convention that the coordinate that is not transformed and kept as reference point is the one associated with the point $x$ with respect to which the $\delta/\delta K(x)$ derivative is taken. With this prescription, dilatation and conformal Ward identities take the following forms for the vertices $\Gamma^{(n,0)}|_{K=0}$:
\begin{equation}\label{dilGamman0}
\begin{aligned}
    &\left[\left(\sum_{i=1}^{n-1}p_i^{\nu}\frac{\partial}{\partial p_i^{\nu}}\right) \right.
    \\
    &-d+nD_{\varphi}\left.\vphantom{\sum_{i=1}^{n-1}}+D_{\varphi}\phi_i\frac{\partial}{\partial \phi_i}\right]\Gamma^{(n,0)}_{a_1\dots a_n}(p_1,\dots,p_{n-1})=
    \\
    &\Tr\Big[\dot{R}_k(q)G\cdot H^{(n,0)}_{\cdot a_1\dots a_n\cdot}(p_1,\dots,p_{n-1})\cdot G\Big],
\end{aligned}
\end{equation}
\begin{equation}\label{confGamman0}
    \begin{aligned}
        \sum_{i=1}^{n-1}\left[\vphantom{\sum_{i=1}^{n}}\right.p_i^{\mu}&\frac{\partial^2}{\partial p_i^{\nu}\partial p_i^{\nu}}-2p_i^{\nu}\frac{\partial^2}{\partial p_i^{\nu}\partial p_i^{\mu}}
        \\
        -&2D_{\varphi}\frac{\partial}{\partial p_i^{\mu}}\left.\vphantom{\sum_{i=1}^{n}}\right]\Gamma^{(n,0)}_{a_1\dots a_n}(p_1,\dots,p_{n-1}) 
        \\
        -2D_{\varphi}\phi_i &\left.\frac{\partial}{\partial r^{\mu}}\Gamma^{(n+1,0)}_{ia_1\dots a_n}(r,p_1,\dots)\right\vert_{r=0}= 
        \\
        -&\int_{q}\dot{R}_k(q)G_{bc}(q)\left(\frac{\partial}{\partial q^{\mu}}+\frac{\partial}{\partial q'^{\mu}}\right)
        \\
        &\left.\left\{H_{ca_1\dots a_n d}^{(n,0)}(q,p_1,\dots,p_{n-1},q')\right\}\right\vert_{q'=-q}G_{db}(q).
    \end{aligned}
\end{equation}

In a similar way, for the vertices $\Gamma^{(n,1)}|_{K=0}$, these identities take the form:

\begin{equation}\label{dilGamman1}
\begin{aligned}
    &\left[\left(\sum_{i=1}^{n}p_i^{\nu}\frac{\partial}{\partial p_i^{\nu}}\right) \right.
    \\
    &-D_{\mathcal{O}}+nD_{\varphi}\left.\vphantom{\sum_{i=1}^{n-1}}+D_{\varphi}\phi_i\frac{\partial}{\partial \phi_i}\right]\Gamma^{(n,1)}_{a_1\dots a_n}(p_1,\dots,p_{n})=
    \\
    &\Tr\Big[\dot{R}_kG\cdot H^{(n,1)}_{\cdot a_1\dots a_n \cdot}(p_1,\dots,p_{n})\cdot G\Big],
\end{aligned}
\end{equation}
\begin{equation}\label{confGamman1}
    \begin{aligned}
\sum_{i=1}^{n}\left[\vphantom{\sum_{i=1}^{n}}\right.p_i^{\mu}&\frac{\partial^2}{\partial p_i^{\nu}\partial p_i^{\nu}}-2p_i^{\nu}\frac{\partial^2}{\partial p_i^{\nu}\partial p_i^{\mu}}
        \\
        +&2(D_{\mathcal{O}}-d)\frac{\partial}{\partial p_i^{\mu}}\left.\vphantom{\sum_{i=1}^{n}}\right]\Gamma^{(n,1)}_{a_1\dots a_n}(p_1,\dots,p_{n}) 
        \\
        -2D_{\varphi}\phi_i \frac{\partial}{\partial r^{\mu}}&\Gamma^{(n+1,1)}_{ia_1\dots a_n}(r,p_1,\dots)\Big\vert_{r=0}=
        -\int_{q}\dot{R}_k(q)G_{bc}(q)\\
        \left(\frac{\partial}{\partial q^{\mu}}+\frac{\partial}{\partial q'^{\mu}}\right)&
        \left.\left\{H_{ca_1\dots a_n d}^{(n,1)}(q,p_1,\dots,p_{n},q')\right\}\right\vert_{q'=-q}G_{db}(q).
    \end{aligned}
\end{equation}

We would like now to highlight the fact that the dilatation Ward identities \eqref{dilGamman0} and \eqref{dilGamman1} are enough to obtain the solution to the fixed point of the FRG flow equations. As a consequence, conformal Ward identities constitute over-restrictive constraints that must be automatically satisfied by the solutions of the dilatation equations.

\subsection{Compatibility between dilation and conformal identities for the $\Gamma_k^{(2,0)}$}\label{CompatGamm2}

The dilatation and conformal Ward identities are, in principle, inequivalent. Since the fixed point of the FRG equation \eqref{WettEqDless} is equivalent to the dilatation Ward identity and conformal symmetry is satisfied at criticality (or at least we assume this to be the case), one expects that the conformal Ward identity could be verified once dilatation and isometries are imposed. However, this is far from trivial, although there exist some sufficient conditions \cite{polchinski1988scale,delamotte2016scale} under which, for certain models, it can be indeed proven \cite{delamotte2016scale,DePolsi2019}. We now recall that for $O(N)$ models the conformal Ward identity for the vertex $\Gamma_{k,n_1n_2}^{(2,0)}(p)$ is not independent of the dilatation Ward identity for the same vertex.

To show this, let us first overlook the aspect of the identities whose dependence we want to demonstrate. The dilation Ward identity for $\Gamma_{k,n_1n_2}^{(2,0)}(p)$ is given by:
\begin{equation}
\label{eq_wardDilG2}
\begin{aligned}
    \left[p^{\mu}\frac{\partial}{\partial p^{\mu}}-d+2D_{\varphi}+\phi_i D_{\varphi}\frac{\partial}{\partial\phi_i}\right]\Gamma^{(2)}_{k,ab}(p)=
    \\ \int_q \dot{R}(q)G_{ij}(q)H^{(2)}_{jabk}(q,p,-q)G_{ki}(q),
\end{aligned}
\end{equation}
while the special conformal transformation's one reads:
\begin{equation}
\label{eq_wardSCTG2}
    \begin{split}
         \Bigg[p^{\mu}\frac{\partial^2}{\partial p^{\nu}\partial p^{\nu}}&-2p^{\nu}\frac{\partial^2}{\partial p^{\nu}\partial p^{\mu}}-2D_{\varphi}\frac{\partial}{\partial p^{\mu}}\Bigg]\Gamma^{(2)}_{k,ab}(p)
         \\ -2D_{\varphi}\phi_i\frac{\partial }{\partial r^{\mu}}&\Gamma^{(3)}_{iab}(r,p)\Big\vert_{r=0} =
        -\int_q \dot{R}(q)G_{ij}(q)
        \\ \Bigg\{\Big(\frac{\partial}{\partial q^{\mu}}+\frac{\partial}{\partial q'^{\mu}}\Big)&H^{(2)}_{jabk}(q,p,q')\Bigg\}\Bigg|_{q'=-q}G_{ki}(q).
    \end{split}
\end{equation}
To begin with, consider the second term in the l.h.s. of \eqref{eq_wardSCTG2}, the term including a derivative of $\Gamma^{(3,0)}_{k,iab}$. After a somewhat straightforward manipulation:
\begin{equation}
\label{eq_manipulacionDerG3}
    \begin{split}
        \left.\phi_i\frac{\partial }{\partial r^{\mu}}\Gamma^{(3)}_{k,iab}(r,p)\right\vert_{r=0} &= \left.\phi_i\frac{\partial }{\partial r^{\mu}}\left[\Gamma^{(3)}_{iba}(r,-p-r)\right]\right\vert_{r=0} \\
        &= \left.\phi_i\frac{\partial }{\partial r^{\mu}}\Gamma^{(3)}_{k,iab}(r,-p-r)\right\vert_{r=0} \\
        &= \left.\phi_i\frac{\partial }{\partial r^{\mu}}\left[\Gamma^{(3)}_{k,iab}(-r,p+r)\right]\right\vert_{r=0} \\
        &= -\left.\phi_i\frac{\partial }{\partial r^{\mu}}\left[\Gamma^{(3)}_{k,iab}(r,p)\right]\right\vert_{r=0} 
        \\&\quad+ \phi_i\frac{\partial }{\partial p^{\mu}}\Gamma^{(3)}_{k,iab}(0,p) 
        \\
        &= \frac{1}{2}\frac{\partial }{\partial p^{\mu}}\phi_i\frac{\partial}{\partial \phi_i}\Gamma^{(2)}_{k,ab}(p),
    \end{split}
\end{equation}
which reduces to
\begin{equation} -2D_{\varphi}\phi_i\frac{\partial}{\partial r^{\mu}}\Gamma^{(3,0)}_{k,iab}(r,p)\left.\vphantom{\frac{\partial}{\partial}}\right\vert_{r=0}=-D_{\varphi}\frac{\partial }{\partial p^{\mu}}\phi_i\frac{\partial}{\partial \phi_i}\Gamma^{(2)}_{k,ab}(p).
\end{equation}
In \eqref{eq_manipulacionDerG3}, it was used firstly, that, due to the $O(N)$ symmetry, any rank-2 tensor is proportional to either $\delta_{ab}$ or $\phi_a\phi_b$ -- both structures symmetric under the exchange of $a$ and $b$ -- and secondly, in the third line, the invariance under parity transformation. A very similar treatment can be given to the r.h.s. of \eqref{eq_wardSCTG2}, and one finds:

\begin{equation}
\label{eq_manipLoop}
    \begin{aligned}
        &\left. \left[ G_{ij}(q) \left\{\left( \frac{\partial}{\partial q^{\mu}} + \frac{\partial}{\partial q'^{\mu}} \right) H^{(2)}_{jabk}(q,p,q')\right\}G_{ki}(q') \right] \right\vert_{q'=-q}
        \\
        &=\left[G_{ij}(q)\left(2 \frac{\partial}{\partial p^{\mu}}H^{(2)}_{jabk}(q,p,q')\right.\right.
        \\
        &\hspace{9mm}-\left.\left.\left.\left(\frac{\partial}{\partial q^{\mu}}+\frac{\partial}{\partial q'^{\mu}}\right)H^{(2)}_{jabk}(q,p,q')\right)G_{ki}(q')\right]\right\vert_{q'=-q} 
        \\
        &=G_{ij}(q)\left(\frac{\partial}{\partial p^{\mu}}H^{(2)}_{jabk}(q,p,-q)\right)G_{ki}(-q).
    \end{aligned}    
\end{equation}
To obtain the first equality, the same reasoning presented in \eqref{eq_manipulacionDerG3} is followed. The second equality, in which the arguments $q$, $q'$ in $H^{(2)}_{ajkb}$ are exchanged holds only due to the presence of the propagators. 

Let us now consider a derivative of Eq.~\eqref{eq_wardDilG2} with respect to $p^{\mu}$:

\begin{equation}
\label{eq_derMomentos}
    \begin{aligned}
        &\frac{\partial}{\partial p^{\mu}}\left[\left(p^{\nu}\frac{\partial}{\partial p^{\nu}}-d+2D_{\varphi}+\phi_i D_{\varphi}\frac{\partial}{\partial \phi_i}\right) \Gamma^{(2)}_{k,ab}(p;\phi) \right]
        \\ 
        &= \left(p^{\nu}\frac{\partial^2}{\partial p^{\nu}\partial p^{\mu}}+\left(2D_{\varphi}-d+1\right)\frac{\partial}{\partial p^{\mu}}\right.
        \\
        &\hspace{25mm}+\left.D_{\varphi}\frac{\partial}{\partial p^{\mu}}\phi_i\frac{\partial}{\partial\phi_i}\right)\Gamma^{(2)}_{k,ab}(p;\phi)
        \\
        &=\left(-p^{\mu}\frac{\partial^2}{\partial p^{\nu}\partial p^{\nu}}+2p^{\nu}\frac{\partial^2}{\partial p^{\nu}\partial p^{\mu}}\right.
        \\
        &\hspace{20mm}+\left.2D_{\varphi}\frac{\partial}{\partial p^{\mu}}+D_{\varphi}\frac{\partial}{\partial p^{\mu}}\phi_i\frac{\partial}{\partial\phi_i}\right)\Gamma^{(2)}_{k,ab}(p; \phi) \\
        &+ \left(p^{\mu}\frac{\partial^2}{\partial p^{\nu}\partial p^{\nu}}-p^{\nu}\frac{\partial^2}{\partial p^{\nu}\partial p^{\mu}}+\left( 1-d\right)\frac{\partial^2}{\partial p^{\mu}}\right)\Gamma^{(2)}_{k,ab}(p;\phi).
   \end{aligned}
\end{equation}

The last line in \eqref{eq_derMomentos} is nothing but the derivative with respect of $p^{\nu}$ of the Ward identity related to rotational invariance of $\Gamma^{(2,0)}_{k,ab}(p)$: 
\begin{equation}
    \bigg(p^\mu\frac{\partial}{\partial p^\nu}-p^\nu\frac{\partial}{\partial p^\mu}\bigg)\Gamma^{(2)}_{k,ab}(p;\phi)=0.
\end{equation}
Therefore, the last line of Eq.~(\ref{eq_derMomentos}) is identically zero. What is left is nothing but the l.h.s. of Eq.~\eqref{eq_wardSCTG2}, after one has incorporated the observation \eqref{eq_manipulacionDerG3}. In the same way, it can be readily seen that the derivative of the r.h.s. of Eq.~\eqref{eq_wardDilG2} with respect to $p^{\mu}$ is simply the r.h.s. of Eq.~\eqref{eq_wardSCTG2}, as seen in \eqref{eq_manipLoop}. Hence, the special conformal Ward identity for the vertex $\Gamma^{(2,0)}_{k,ab}(p)$ holds true once translation, rotation, dilation and parity symmetries are enforced. It is worth noting that for $\Gamma^{(n)}$ vertices with $n>2$ a simple relation like the one just mentioned is not known.

\section{Conformal constraints derivative expansion results}\label{secRes}

In this section we study the conformal symmetry group upon the implementation of the DE and present our results. Firstly, we establish the condition that must be satisfied if special conformal transformations symmetry is to hold at criticality, but that is ultimately violated for $O(N)$ models due to the approximation being implemented, i.e. the truncation in the effective action. We first consider the more physical and interesting models $O(1)$, $O(2)$, $O(3)$ and $O(4)$. Then we consider larger values of $N$, namely the $O(10)$, $O(20)$, and $O(100)$ models, and compare with predictions coming from the large $N$ expansion \cite{DAttanasio:1997yph}. It is then evidenced a qualitative change in the behavior for the critical exponent $\omega$ with respect to small values of $N$. We therefore consider a third study case which consist in the $O(5)$ model where a pathology is observed. Finally, we consider the extension of the $O(N)$ models to the non-positive values of $N$ which are related to non-unitary models and study the cases of $N=0$ and $N=-2$.

\subsection{The derivative expansion in the presence of composite operators and the special conformal constraint}

In this section we employ the derivative expansion to find solutions to the dilatation Ward identity corresponding to the critical fixed point of $O(N)$ models for finite values of $N$ and analyze the behavior of the special conformal Ward identity. For this purpose we will generalize the strategy initially presented for the case $N=1$ in Ref.~\cite{Delamotte2024}. In a succinct manner, the procedure is as follows: we consider the \textit{ansatz} for the effective action given in equation \eqref{ansatz} and insert it into the Ward identity for dilatations which, as seen in the previous sections, amounts to finding a fixed point solution to the FRG flow equation \eqref{WettEqDless}. From this, we can extract a set of independent equations that allows us to solve all functions of the \textit{ansatz}.

Evaluating at $K=0$ the equations corresponding to the vertices $\Gamma_{k,a}^{(1,0)}$ and $\Gamma_{k,ab}^{(2,0)}$ yields the fixed point functions $U_0$, $Z_0$ and $Y_0$ (which are denoted with a star, i.e. $U_0^*$, $Z_0^*$ and $Y_0^*$) following the procedure described in App.~\ref{Ap:numerics}. The corresponding equations are given in the supplemental material. A remark is now necessary: the equations for the functions $U_0$, $Z_0$ and $Y_0$ obtained from a $\Gamma_{k}^{(n,0)}$ vertex are independent from those coming from the vertex $\Gamma_{k}^{(n,1)}$.\footnote{It is a general property that the equation for $\Gamma_{k}^{(n,m)}$ is independent of the vertices $\Gamma_{k}^{(n',m')}$ given that $m'>m$, see \cite{Delamotte2024}.} This implies that there is no feedback from functions $U_1$, $Z_1$, $Y_1$ and $\Upsilon$ on $U_0$, $Z_0$ and $Y_0$.

Once the fixed point functions, in the absence of the source for composite operators, are obtained, we \textit{switch on} the source $K$ in order to simultaneously determine the fixed point solution for the functions $U_1$, $Z_1$ and $Y_1$ and find the value of $D_\mathcal{O}$. It so happens that these are independent of the function $\Upsilon$ for structural reasons, see \cite{Delamotte2024}. While the solutions for $U_0^*$, $Z_0^*$ and $Y_0^*$ are unique for each fixed point of the theory, there is an infinite discrete set of solutions for $U_1^*$, $Z_1^*$ and $Y_1^*$ (and $D_\mathcal{O}$) once the fixed point has been chosen: each corresponds to a particular scaling operator $\mathcal{O}$ (or, in equivalent terms, to a particular eigenperturbation around a given fixed point. In this work, we will concentrate on the scaling operators associated with the critical exponents $\nu$, which is the relevant one for the critical point, and the one associated with the critical exponent $\omega$, corresponding to the first correction to scaling. It should be noted that, because the modified conformal identity for $\Gamma_{k,ab}^{(2,0)}$ is automatically satisfied at the fixed point for $O(N)$ models as shown in section \ref{CompatGamm2}, it is not possible to use the present strategy for the purpose of optimizing the exponent $\eta$ at order $\mathcal{O}(\partial^2)$ of the DE. It is possible to obtain an optimization criterion for the exponent at the next order of this expansion \cite{Balog2020}. 

The last step of the algorithm consists in determining the function $\Upsilon$, using that its dilatation Ward identity is a linear differential equation and that all the other functions of the \textit{ansatz} have already been fixed. It is worth emphasizing that the function $\Upsilon$ does not affect the eigenperturbations for uniform values of $K$ although it is part of the general \textit{ansatz} expression, because it expresses the behaviour of the scaling operator in the presence of {\it inhomogeneous} sources $K(x)$. As we discuss below, this function plays an important role in the conformal constraint precisely for this reason. The equations that define functions $U_1$, $Z_1$, $Y_1$ and $\Upsilon$ can be found in the supplemental material.

Before proceeding with the discussion of the conformal constraint, it will prove useful to analyze the behavior of the fixed point equations for the perturbation functions for large values of $\rho$. A property inherited directly from the one-loop structure of the FRG evolution equation for $\Gamma_k$ Eq.~\eqref{eq_manipLoop} and the subsequent change to dimensionless variables is that fixed point equations have a \textit{dimensional} part and a loop contribution. In the large $\rho$ region the loop contribution to the equations is negligible in comparison with the \textit{dimensional} part, as can be readily checked. This implies, as a direct consequence, that the behavior of functions $Z_1$, $Y_1$ and $\Upsilon$ at large values of $\rho$ is the following:

\begin{equation}\label{largeFieldScaling}
    \begin{split}
	Z_1(\rho)&\underset{\rho\gg 1}{\sim} A_{Z_1}\rho^{(D_{\mathcal{O}}-d-\eta)/(d-2+\eta)},\\
	Y_1(\rho)&\underset{\rho\gg 1}{\sim} A_{Y_1}\rho^{(D_{\mathcal{O}}-2d+2-2\eta)/(d-2+\eta)},\\
	\Upsilon(\rho)&\underset{\rho\gg 1}{\sim} A_{\Upsilon}\rho^{(D_{\mathcal{O}}-2)/(d-2+\eta)},
    \end{split}
\end{equation}
where $A_{Y_1}$, $A_{Z_1}$ and $A_{\Upsilon}$ are constants which are fixed up to a normalization. These constant are, in fact, determined by the behavior at smaller values of $\rho$, regime in which the loop term intervenes.

Let us now consider special conformal symmetry. When inserting the \textit{ansatz} into the conformal Ward identity for the vertex $\Gamma_{k,\vec{a}}^{(1,1)}$ one arrives at a relation which has the following form:
\begin{equation}\label{confConstDE}
\begin{split}
C_L(\rho)\equiv2\Big(&(d-2)\Upsilon'(\rho)+4\rho D_\varphi\Upsilon''(\rho)\\&-D_\varphi \big(Z_1(\rho)+\rho Y_1(\rho)\big)\Big)=C_R(\rho),
\end{split}
\end{equation}
where the term $C_R(\rho)$ corresponds to the loop contribution which is given in the supplemental material. As discussed previously, all functions are already determined by means of the modified dilatation Ward identities. This implies that equation \eqref{confConstDE} is an over-constraint which will be, generically, violated once approximations are performed. In particular, this applies also to the large field behavior for each function in the ansatz which is given in \eqref{largeFieldScaling}. Now, at large fields, as was the case when discussing the asymptotic behavior of $Z_1$, $Y_1$ and $\Upsilon$, the constraint given in equation \eqref{confConstDE} is dominated only by its \textit{left hand side}. However, when substituting the large field behavior \eqref{largeFieldScaling}, the whole expression scales at large fields as $\rho^{(D_{\mathcal{O}}-d-\eta)/(d-2+\eta)}$. Unless the following relationship holds
\begin{equation}\label{exactConfConst}
    A_{\Upsilon}=2D_\varphi^2\frac{A_{Z_1}+A_{Y_1}}{(D_{\mathcal{O}}-2)(d+2D_{\mathcal{O}}-4D_\varphi-6)},
\end{equation}
the large field behavior of the conformal constraint will be suppressed for relevant perturbations and will grow as a power law for irrelevant perturbations. This constant prefactors, as previously explained are determined by the small and medium field regime of the modified dilatation Ward identities. In the exact theory, equation \eqref{exactConfConst} would be satisfied, but it is not in view of the approximations implemented, namely the DE.\footnote{The large field behavior grows as a power law for irrelevant perturbations that satisfy $D_{\mathcal{O}}>d+\eta$, which in practice is all irrelevant perturbations for the $O(N)$ model.} To overcome this obstacle and to be able to evaluate the conformal constraint \eqref{confConstDE} without the large field bias, we normalize it in the following manner:
\begin{equation}\label{confConstDENorm}
	f(\rho;\alpha)=\left[C_L(\rho)-C_R(\rho)\right]\left(1+\frac{\rho}{\rho_0}\right)^{\frac{d+\eta-D_{\mathcal{O}}}{d-2+\eta}},
\end{equation}
where we have also introduced an intermediate scale $\rho_0$ in order to avoid spoiling the small field behavior. In general, we choose $\rho_0$ as the minimum of the potential.\footnote{Given the fact that for the case $N=-2$ the minimum is located at $\rho=0$, we use a different criteria for $\rho_0$ in \eqref{confConstDENorm} for this model. We choose, for this case, the first local extreme of the $Y_0$ function -- since $Z_0$ does neither have a extreme in the studied region.} We remark that this value of $\rho_0$ is an $\alpha$ dependent parameter which also levels the comparison of the conformal constraint for different values of $\alpha$.

One way of exploiting the normalized conformal constraint \eqref{confConstDENorm} is to use it in order to fix spurious parameters of the scheme by means of what we called the PMC criterion. The idea is to evaluate the $f$ function at the fixed point and search for which values of the scheme parameter $\alpha$, say $\alpha_{PMC}$, the function is closer to zero with regards to some measure. Since for each value of $\alpha$ there's a whole function of $\rho$ there are many possible implementations of the PMC idea. In this and previous works \cite{Balog2020,Delamotte2024}, we chose to fix $\alpha_{PMC}$ as the value of $\alpha$ for which the function $f(\rho=0,\alpha)$ is closer to zero. It is worth emphasizing that different implementations of the PMC criterion lead to results which are compatible (that is, within error bars).

All the analyzed values of $N$ where studied previously in detail, with regards to critical exponents and the DE, in the study presented in \cite{DePolsi2020}. However, even though it was performed at order $\mathcal{O}(\partial^4)$ of the DE (or even at order $\mathcal{O}(\partial^6)$ for $N=1$ \cite{Balog:2019rrg}), it was done using the strict version of the DE expansion, while, as already mentioned, in this work we will consider the full version of the scheme.\footnote{The full version of the DE was considered at order $\mathcal{O}(\partial^4)$ in \cite{Peli:2020yiz}, although a field expansion was considered on top of it.} This provides us with the opportunity to compare both truncations at order $\mathcal{O}(\partial^2)$ of the DE.  As can be seen from the Tables~Tables~\ref{tab:N=1} to \ref{tab:N=-2}, for all exponents and for all values of $N$, the PMS criterion gives almost identical results with differences well below the error bars for both implementations.  This first comparison between the \textit{full} and \textit{strict} versions of the DE seems to demonstrate their equivalence (at least up to the precision level of the order $\mathcal{O}\left(\partial^2\right)$ of the DE) as one could \textit{a priori} expect, since they differ by higher order terms. A similar comment applies when using the PMC criterion, except that for the critical exponent $\eta$ where we do not have at our disposal a constraint because, as explained before, there is no conformal constraint for the leading odd operator. Consequently, we cannot use PMC to fix a criterion at this order of the DE for that exponent.

In what follows, we restrict our analysis to two widely utilized families of such regulating functions, 
\begin{equation}\label{regprofiles}
\begin{split}
    E_k(q^2)&=\alpha Z_k k^2 \text{e}^{-q^2/k^2} 
    \\
    W_k(q^2)&=\alpha Z_k k^2 \frac{q^2/k^2}{\text{e}^{q^2/k^2-1}}
\end{split}
\end{equation}
which we denominate, respectively, exponential and Wetterich \cite{Wetterich:1992yh} regulators.

\subsection{Results for $N=1, 2, 3$, and $4$}

Let us begin by considering the physically interesting $\mathbb{Z}_2$ (equivalent to $O(1)$), $O(2)$, $O(3)$ and $O(4)$ models. Since we only seek here to explore the plausibility of implementing the conjectured conformal symmetry to extract physical predictions, we present a comparison with previous results within DE and the most precise result available in the literature.

\begin{figure*}[htbp]
    \centering
    \begin{minipage}{0.46\textwidth}
        \centering
        \includegraphics[width=\textwidth]{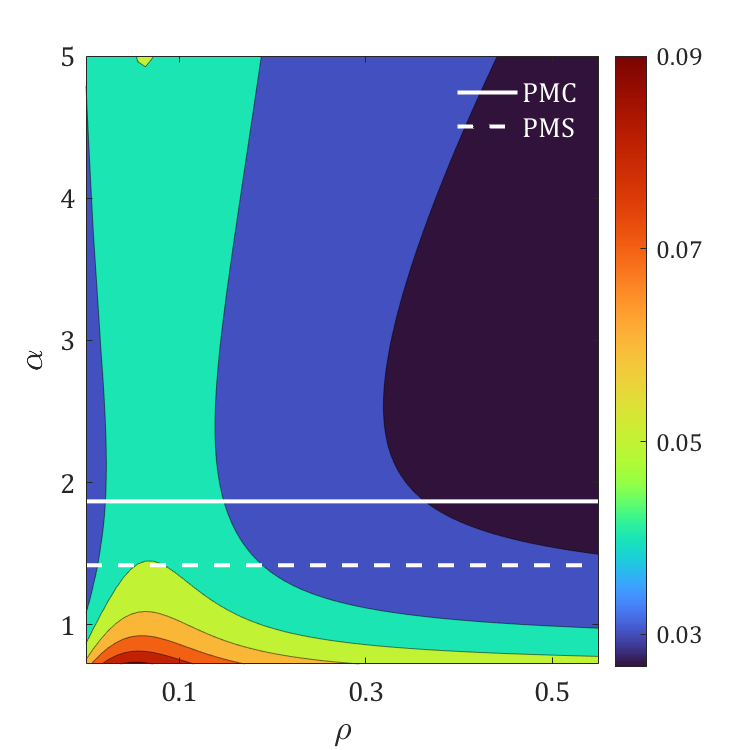}
        \caption{Function $f(\rho,\alpha)$ corresponding to the relevant perturbation $\nu$ for the $O(2)$ model. The continuous line indicates the $\alpha_{\text{PMC}}$ value, while the dashed one indicates the $\alpha_{\text{PMS}}$ one. This figure corresponds to calculations performed with the exponential regulator given in \eqref{regprofiles}.}
        \label{FigNu2}
    \end{minipage}
    \hfill
    \begin{minipage}{0.46\textwidth}
        \centering
        \includegraphics[width=\textwidth]{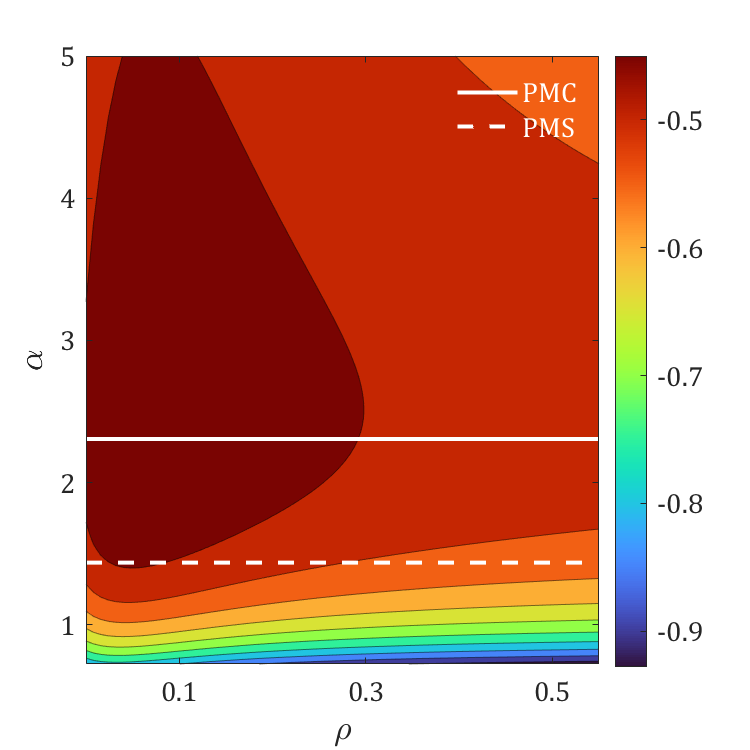}
        \caption{Function $f(\rho,\alpha)$ corresponding to the irrelevant perturbation $\omega$ for the $O(2)$ model. The continuous line indicates the $\alpha_{\text{PMC}}$ value, while the dashed one indicates the $\alpha_{\text{PMS}}$ one. This figure corresponds to calculations performed with the exponential regulator given in \eqref{regprofiles}.}
        \label{FigOmega2}
    \end{minipage}
    
    \vspace{0.5cm} 
    \begin{minipage}{0.46\textwidth}
        \centering
        \includegraphics[width=\textwidth]{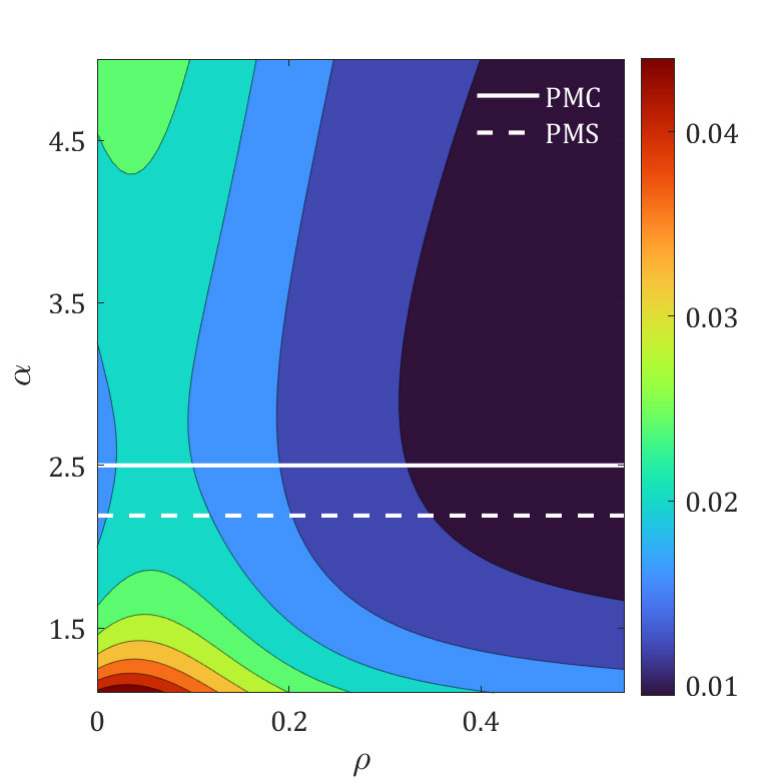}
        \caption{Function $f(\rho,\alpha)$ corresponding to the relevant perturbation $\nu$ for the $O(4)$ model. The continuous line indicates the $\alpha_{\text{PMC}}$ value, while the dashed one indicates the $\alpha_{\text{PMS}}$ one. This figure corresponds to calculations performed with the exponential regulator given in \eqref{regprofiles}.}
        \label{FigNu4}
    \end{minipage}
    \hfill
    \begin{minipage}{0.46\textwidth}
        \centering
        \includegraphics[width=\textwidth]{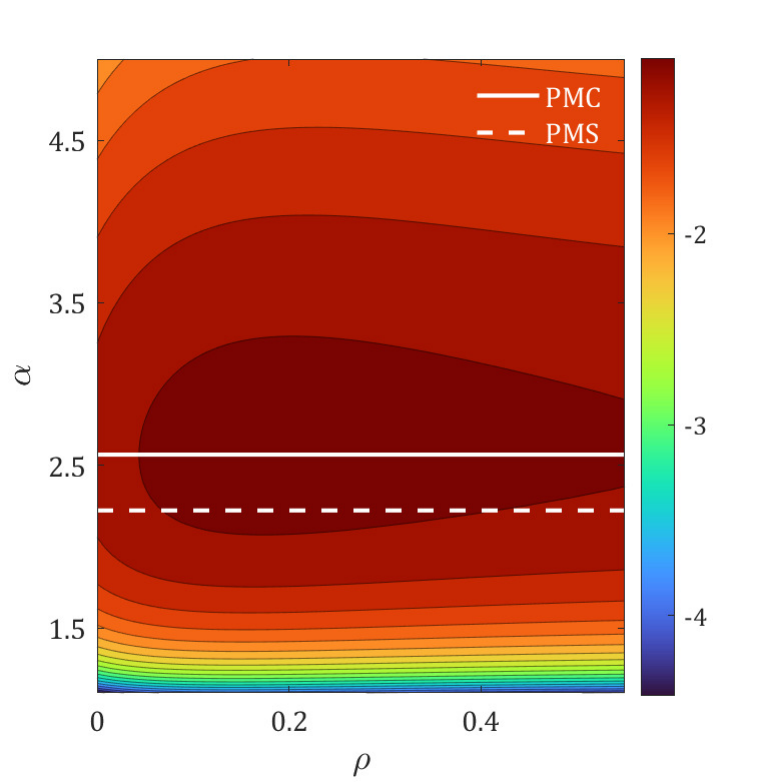}
        \caption{Function $f(\rho,\alpha)$ corresponding to the irrelevant perturbation $\omega$ for the $O(4)$ model. The continuous line indicates the $\alpha_{\text{PMC}}$ value, while the dashed one indicates the $\alpha_{\text{PMS}}$ one. This figure corresponds to calculations performed with the exponential regulator given in \eqref{regprofiles}.}
        \label{FigOmega4}
    \end{minipage}
\end{figure*}

We now discuss the behavior of the $f(\rho;\alpha)$ function given in equation \eqref{confConstDENorm} for small values of $N$. All models considered here show a qualitatively similar behavior, for both the relevant perturbation $\nu$ and least irrelevant one, $\omega$ (as can be seen in Figs.~\ref{FigNu2} to \ref{FigOmega4} for the $O(2)$ and $O(4)$ models for example). In all these cases, there was a region of the parameter $\alpha$ of the regulator family in which the conformal constraint breaking was minimized. As can be seen in the mentioned figures, the corresponding $\alpha_{\text{PMC}}$ differed only by a factor of order one to those coming from the PMS criterion. Considering the weak dependence of the critical exponents in the $\alpha$ parameter over such variations, our studies gives to very similar results for $\nu$ and $\omega$ by imposing either the PMC or PMS criteria. Our results, compared to those from the strict version of the DE and the most precise available estimation are presented in Tables~\ref{tab:N=1} to \ref{tab:N=4}. For all considered cases, and for both $\nu$ and $\omega$, the predictions using either PMS or PMC criteria coincide within error bars, sometimes even having identical central values. 

\begin{table}[!ht]
    \centering
    \begin{tabular}{l l l l}
    \toprule[1pt]\midrule[0.3pt]
          & $\nu$   & $\eta$ & $\omega$ \\
    \midrule     
    $\mathcal{O}\left(\partial^2\right)_{\text{fPMC}}$   & 0.6308(27)   & & 0.849(49) \\
    $\mathcal{O}\left(\partial^2\right)_{\text{fPMS}}$ & 0.6309(27)  & 0.0387(55) & 0.848(49)  \\
    \midrule 
    $\mathcal{O}\left(\partial^2\right)_{\text{sPMS}}$   &  0.6308(27) & 0.0387(55) & 0.870(55)  \\
    $\mathcal{O}\left(\partial^4\right)_{\text{sPMS}}$   &  0.62989(25) & 0.0362(12) & 0.832(14)  \\
    $\mathcal{O}\left(\partial^6\right)_{\text{sPMS}}$   &  0.63012(16) & 0.0361(11) & \\
    \midrule
    CB  &  0.629971(4) &  0.0362978(20) &0.82968(23)  \\
    \midrule[0.3pt]\bottomrule[1pt]
    \end{tabular}
    \caption{Results obtained, with their corresponding error bars, for the $\mathbb{Z}_2$ model (corresponding to $N=1$). We report the predicted exponents through both the PMC and PMS criteria with the full version of the DE at order $\mathcal{O}\left(\partial^2\right)$. We also include the results, for various orders of the DE, obtained with the strict version from \cite{Balog2019,DePolsi2020}. As a comparison benchmark, we include as well the most precise available results from the CB program (\cite{Kos:2014bka} for $\nu$ and $\eta$ and \cite{Simmons-Duffin:2016wlq} for $\omega$).}
    \label{tab:N=1}
\end{table}

\begin{table}[!ht]
    \centering
    \begin{tabular}{l l l l}
    \toprule[1pt]\midrule[0.3pt]
          & $\nu$   & $\eta$ & $\omega$  \\
    \midrule     
    $\mathcal{O}\left(\partial^2\right)_{\text{fPMC}}$   & 0.6726(52)   &  & 0.786(29)\\
    $\mathcal{O}\left(\partial^2\right)_{\text{fPMS}}$ & 0.6727(52)  & 0.0408(58) &  0.787(30)\\
    \midrule 
    $\mathcal{O}\left(\partial^2\right)_{\text{sPMS}}$   &  0.6725(52)  & 0.0410(59) & 0.798(34)\\
    $\mathcal{O}\left(\partial^4\right)_{\text{sPMS}}$   &  0.6716(6) & 0.0380(13) & 0.791(8)\\
    \midrule
    MC (2019)  & 0.67169(7)   & 0.03810(8) &  0.789(4)\\
    CB & 0.6718(1) & 0.03818(4) & 0.794(8) \\
    \midrule[0.3pt]\bottomrule[1pt]
    \end{tabular}
    \caption{Results obtained, with their corresponding error bars, for the $O(2)$ model. We report the predicted exponents through both the PMC and PMS criteria with the full version of the DE at order $\mathcal{O}\left(\partial^2\right)$. We also include the results obtained with the strict version from \cite{DePolsi2020}. As a comparison benchmark, we include as well the most precise available results from MC calculations \cite{Hasenbusch:2019jkj} and \textit{conformal bootstrap} restrictions \cite{chester2019carving}.}
    \label{tab:N=2}
\end{table}

\begin{table}[!ht]
    \centering
    \begin{tabular}{l l l l}
    \toprule[1pt]\midrule[0.3pt]
          & $\nu$   & $\eta$ & $\omega$  \\
    \midrule     
    $\mathcal{O}\left(\partial^2\right)_{\text{fPMC}}$   & 0.7125(71)   & & 0.744(26)\\
    $\mathcal{O}\left(\partial^2\right)_{\text{fPMS}}$ & 0.7126(71)  & 0.0405(58) & 0.746(26)\\
    \midrule 
    $\mathcal{O}\left(\partial^2\right)_{\text{sPMS}}$   &  0.7125(71)   &  0.0408(58) & 0.754(34)\\
    $\mathcal{O}\left(\partial^4\right)_{\text{sPMS}}$   &  0.7114(9) & 0.0376(13) & 0.769(11)\\
    \midrule
    $\epsilon-$exp, $\epsilon^6$    &  0.7059(20)   & 0.0378(5) & 0.795(7)\\
    MC + High T & 0.7112(5)    & 0.0375(5) &  \\
    \midrule[0.3pt]\bottomrule[1pt]
    \end{tabular}
    \caption{Results obtained, with their corresponding error bars, for the $O(3)$ model. We report the predicted exponents through both the PMC and PMS criteria with the full version of the DE at order $\mathcal{O}\left(\partial^2\right)$. We also include the results obtained with the strict version from \cite{DePolsi2020}. As a comparison benchmark, we include as well the most precise available results from the $\epsilon-$expansion to $\epsilon^6$ order \cite{Kompaniets:2017yct} for $\omega$ and combined MC calculations with high temperature expansion \cite{Campostrini_2002} for $\omega$.}
    \label{tab:N=3}
\end{table}

\begin{table}[!ht]
    \centering
    \begin{tabular}{l l l l}
    \toprule[1pt]\midrule[0.3pt]
          & $\nu$  & $\eta$ & $\omega$  \\
    \midrule     
    $\mathcal{O}\left(\partial^2\right)_{\text{fPMC}}$   & 0.749(8)   & & 0.723(26)\\
    $\mathcal{O}\left(\partial^2\right)_{\text{fPMS}}$ & 0.749(8)  & 0.0387(55) & 0.723(26)\\
    \midrule 
    $\mathcal{O}\left(\partial^2\right)_{\text{sPMS}}$   &  0.749(8) & 0.0389(56) & 0.731(34)\\
    $\mathcal{O}\left(\partial^4\right)_{\text{sPMS}}$   &  0.7478(9) & 0.0360(12)  & 0.761(12)\\
    \midrule
    MC  & 0.74817(20)  & 0.0360(4)  & 0.755(5) \\
    \midrule[0.3pt]\bottomrule[1pt]
    \end{tabular}
    \caption{Results obtained, with their corresponding error bars, for the $O(4)$ model. We report the predicted exponents through both the PMC and PMS criteria with the full version of the DE at order $\mathcal{O}\left(\partial^2\right)$. We also include the results obtained with the strict version from \cite{DePolsi2020}. As a comparison benchmark, we include as well the most precise available results from MC calculations \cite{Hasenbusch_2022}.}
    \label{tab:N=4}
\end{table}

\subsection{Results for larger values of $N$}

Let us now turn our attention to large values of $N$ -- namely the $O(10)$, $O(20)$ and $O(100)$ models. Even though lacking physical realizations, they are of theoretical interest as the exact solution for the large $N$ limit has been calculated decades ago and so have been the leading corrections in a $1/N$ expansion \cite{Okabe78,Vasil'ev1982,Broadhurst:1996ur}. Therefore, our results can be compared to the exact asymptotic critical exponents. These expressions for $\eta$, $\nu$ and $\omega$ read:
\begin{equation}
\label{eq_asymptoticNuOmega}
    \begin{split}
\eta &= \frac{8}{3\pi^2}\frac{1}{N}-\frac{512}{27\pi^4}\frac{1}{N^2}
-\frac{8}{27\pi^6}\frac{1}{N^3}\\
&\times \Bigg[
\frac{797}{18}-\zeta(2)\Big(27 \log(2)-\frac{61}{4}\Big)+\zeta(3)\frac{189}{4}\Bigg]+\mathcal{O}\left(\frac{1}{N^4}\right) \\
        \nu &= 1-\frac{32}{3\pi^2}\frac{1}{N}-\frac{32}{27\pi^4}(27\pi^2-112)\frac{1}{N^2}+\mathcal{O}\left(\frac{1}{N^3}\right) \\
        \omega &= 1-\frac{64}{3\pi^2}\frac{1}{N}+\frac{128}{9\pi^4}\left(\frac{104}{3}-\frac{9\pi^2}{2}\right)\frac{1}{N^2}+\mathcal{O}\left(\frac{1}{N^3}\right).
    \end{split}
\end{equation}

In Tables~\ref{tab:N=10} to \ref{tab:N=100} we present our results, compared to the ones from the strict version and those of the large $N$ expansion. For these, we employ \eqref{eq_asymptoticNuOmega} to estimate the central values, and the difference between the two last calculated orders in the expansion to approximate their error bars -- although this may be too pessimistic a estimation, we prefer to be conservative. 

\begin{table}[!ht]
    \centering
    \begin{tabular}{l l l l}
    \toprule[1pt]\midrule[0.3pt]
          & $\nu$ & $\eta$  & $\omega$  \\
    \midrule     
    $\mathcal{O}\left(\partial^2\right)_{\text{fPMC}}$   & 0.879(10)   & & 0.782(26)\\
    $\mathcal{O}\left(\partial^2\right)_{\text{fPMS}}$ & 0.879(10)  & 0.0236(34) & 0.783(26)\\
    \midrule 
    $\mathcal{O}\left(\partial^2\right)_{\text{sPMS}}$   &  0.877(11) & 0.0240(34) & 0.788(26)\\
    $\mathcal{O}\left(\partial^4\right)_{\text{sPMS}}$   &  0.8776(10) & 0.0231(6)  & 0.807(7)\\
    \midrule
    Large N  &  0.87(2) & 0.023(2) & 0.77(1)\\
    \midrule[0.3pt]\bottomrule[1pt]
    \end{tabular}
    \caption{Results obtained, with their corresponding error bars, for the $O(10)$ model. We report the predicted exponents through both the PMC and PMS criteria with the full version of the DE at order $\mathcal{O}\left(\partial^2\right)$. We also include the results obtained with the strict version from \cite{DePolsi2020} and the results coming from the $1/N$ expansion to $(1/N^3)$ order.}
    \label{tab:N=10}
\end{table}

\begin{table}[!ht]
    \centering
    \begin{tabular}{l l l l}
    \toprule[1pt]\midrule[0.3pt]
          & $\nu$ & $\eta$  & $\omega$  \\
    \midrule     
    $\mathcal{O}\left(\partial^2\right)_{\text{fPMC}}$   & 0.9429(46)   & &0.886(14)\\
     $\mathcal{O}\left(\partial^2\right)_{\text{fPMS}}$ &0.9428(46) & 0.0126(18) & 0.886(14)\\
    \midrule 
    $\mathcal{O}\left(\partial^2\right)_{\text{sPMS}}$   &   0.9414(49) & 0.0130(19)   & 0.887(14)\\
    $\mathcal{O}\left(\partial^4\right)_{\text{sPMS}}$   &  0.9409(6) & 0.0129(3) & 0.887(2)\\
    \midrule
    Large N    & 0.941(5) & 0.0128(2) & 0.888(3) \\
    \midrule[0.3pt]\bottomrule[1pt]
    \end{tabular}
    \caption{Results obtained, with their corresponding error bars, for the $O(20)$ model. We report the predicted exponents through both the PMC and PMS criteria with the full version of the DE at order $\mathcal{O}\left(\partial^2\right)$. We also include the results obtained with the strict version from \cite{DePolsi2020} and the results coming from the $1/N$ expansion to $(1/N^3)$ order.}
    \label{tab:N=20}
\end{table}

\begin{table}[!ht]
    \centering
    \begin{tabular}{l l l l}
    \toprule[1pt]\midrule[0.3pt]
          & $\nu$ & $\eta$ & $\omega$  \\
    \midrule     
    $\mathcal{O}\left(\partial^2\right)_{\text{fPMC}}$   & 0.98939(82) &  & 0.9771(29)\\
     $\mathcal{O}\left(\partial^2\right)_{\text{fPMS}}$ & 0.98939(82) &  0.00259(37) & 0.9782(32)\\
    \midrule 
     $\mathcal{O}\left(\partial^2\right)_{\text{sPMS}}$   & 0.9892(11) & 0.00257(37) & 0.9782(26)\\
     $\mathcal{O}\left(\partial^4\right)_{\text{sPMS}}$   & 0.9888(2) & 0.00268(4) & 0.9770(8)\\
    \midrule
    Large N & 0.9890(2) & 0.002681(1)& 0.9782(2)\\
    \midrule[0.3pt]\bottomrule[1pt]
    \end{tabular}
    \caption{Results obtained, with their corresponding error bars, for the $O(100)$ model. We report the predicted exponents through both the PMC and PMS criteria with the full version of the DE at order $\mathcal{O}\left(\partial^2\right)$. We also include the results obtained with the strict version from \cite{DePolsi2020} and the results coming from the $1/N$ expansion to $(1/N^3)$ order.}
    \label{tab:N=100}
\end{table}

As can be readily seen, for these large values of $N$ also the PMC and PMS criteria lead again to equivalent predictions with identical error bars and very similar central values for both $\nu$ and $\omega$. Furthermore, our results are in agreement with the large $N$ expansion. This allows us to suppose that as $N$ is increased, the DE converges nicely to the exact solution.

\subsection{The $O(5)$ model: a pathology in the conformal constraint}

Both for $N\leq 4$ and for large values of $N$, the comparison of the PMS and PMC criteria works very well for the critical exponent considered, $\nu$ or $\omega$. However, a peculiar situation arises for the $N=5$ case, as we discuss now.

If one considers the leading scaling operator associated with the critical exponent $\nu$, one arrives, as before, at similar curves and conclusions than for other values of $N$. However, the situation changes drastically when considering the first correction to scaling associated with critical exponent $\omega$. 

\begin{figure}[h!]
	\centering
	\includegraphics[width=\columnwidth]{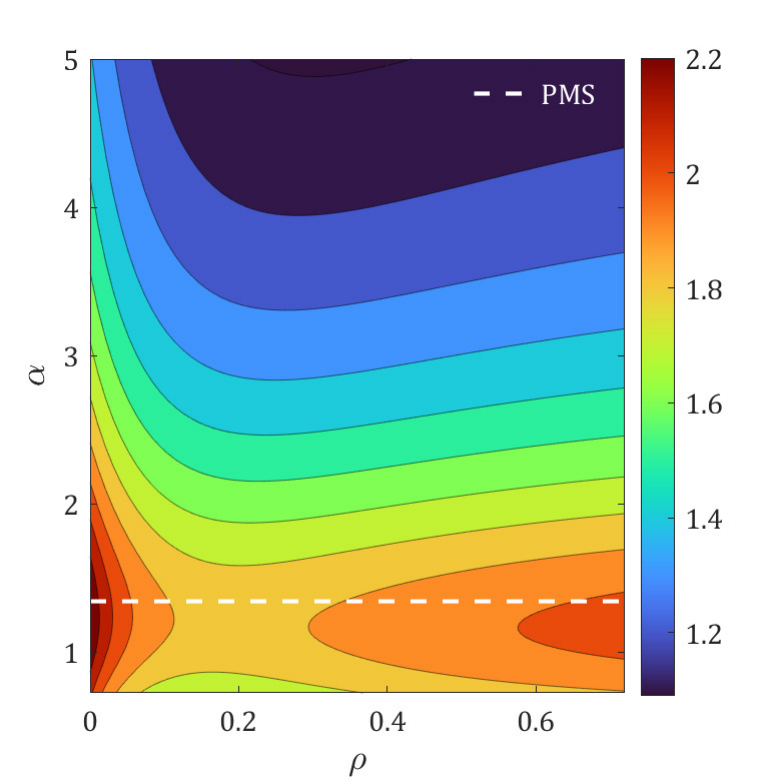}
	\caption{Function $f(\rho,\alpha)$ corresponding to the irrelevant perturbation $\omega$ for the $O(5)$ model. The dashed line indicates the $\alpha_{\text{PMS}}$ value. Notice that the function $|f(\rho,\alpha)|$ presents a maximum, rather than a minimum, thus the absence of the $\alpha_{\text{PMC}}$ value. This figure corresponds to calculations performed with the exponential regulator given in \eqref{regprofiles}.}
	\label{FigConfOmegN5}
\end{figure}

\begin{figure}[h!]
	\centering
	\includegraphics[width=\columnwidth]{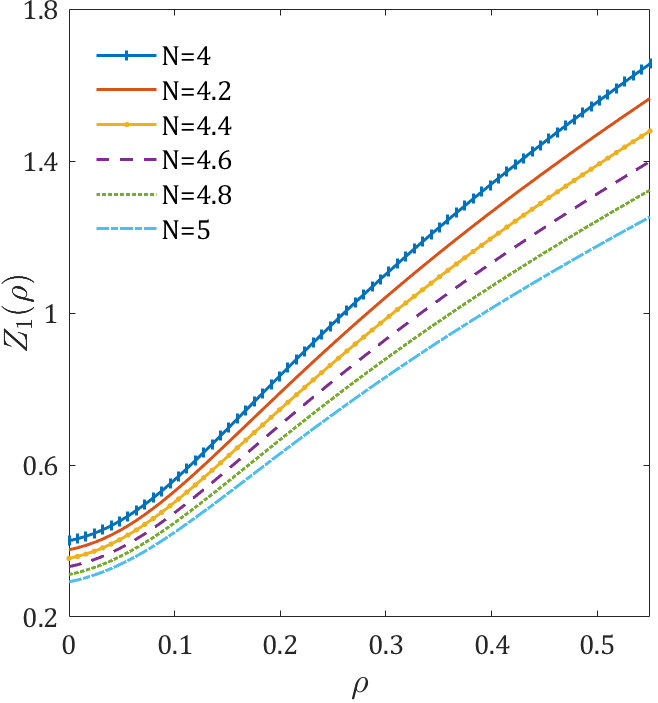}
	\caption{Evolution of the profile $Z_1(\rho)$ as the $N$ parameter is varied from $N=4$ to $N=5$, for the least irrelevant perturbation. The curves correspond to $\alpha=1$, using the exponential regulator given in \eqref{regprofiles}. As can be readily seen, the successive profiles evolve smoothly through the (un-physical) non-integer N values.}
	\label{FigZ1N5}
\end{figure}

\begin{figure}[htpb]
    \centering
    \includegraphics[width=\columnwidth]{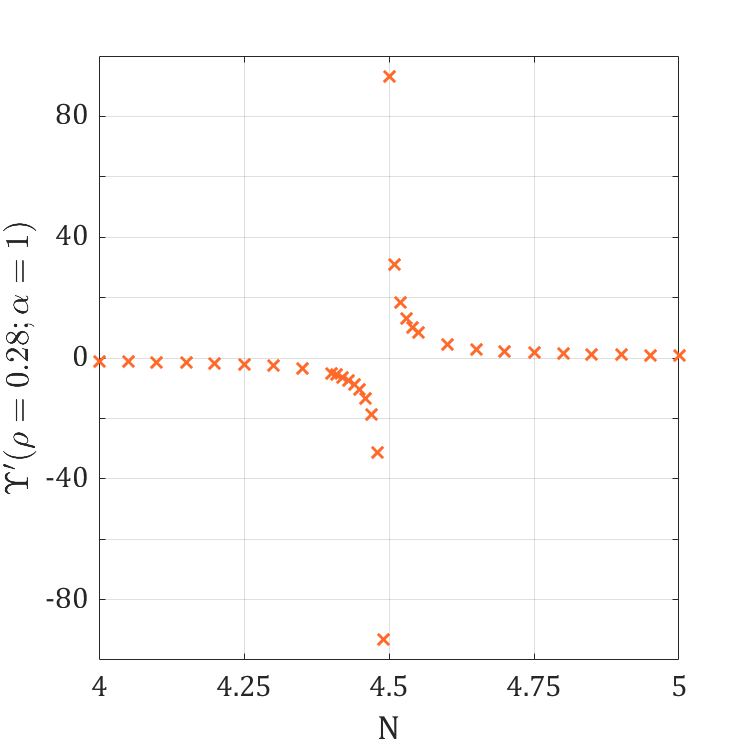}
    \caption{Evolution of the \textit{ansatz} function $\Upsilon'(\rho)$ for a fixed value of $\rho$ and regulator parameter $\alpha=1$, while changing the $N$ parameter characterizing the $O(N)$ symmetry from $N=4$ to $N=5$. The value of $\rho$ was chosen to be the middle of the considered box. The qualitative behavior is independent of this choice, and is exhibited for every value of $\rho$ taken into consideration. As can be seen, the function does not evolve smoothly between the two physical values $N=4$ and $N=5$, but instead shows a non-analyticity near $N\approx 4.5$. The value of $N$ where the divergent behavior takes place depends on the regulator -- for this plot the exponential one was employed -- and the $\alpha$ parameter, although seems to be independent of the $\rho$ value. Further increasing $N$ leads to a smooth behavior for all considered values of $N$}
    \label{FigUpsN5}
\end{figure}

\begin{figure}[h!]
	\centering
	\includegraphics[width=\columnwidth]{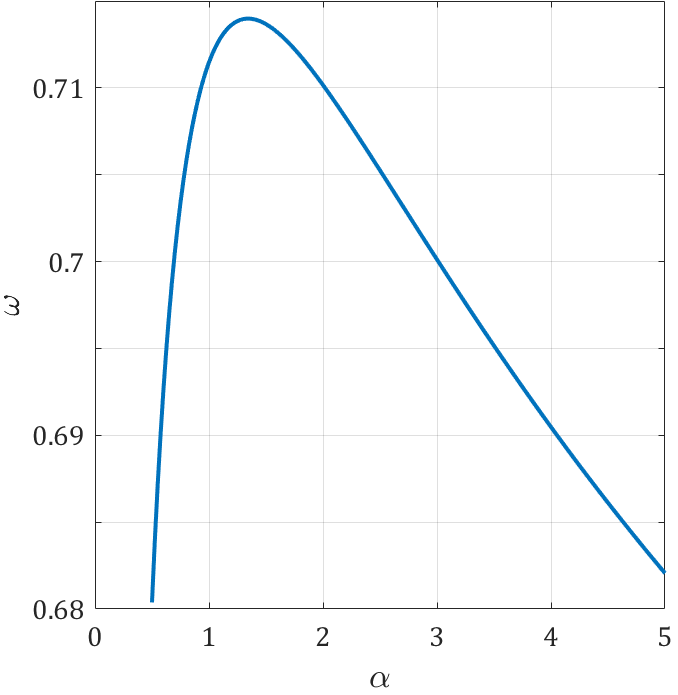}
	\caption{Behaviour of the critical exponent $\omega$, related to the first correction to scaling, as a function of the regulator parameter $\alpha$, for the $O(5)$ model. The represented curve corresponds to the exponential regulator given in \eqref{regprofiles}.}
	\label{FigOmvsAlphaN5}
\end{figure}

In Fig.~\ref{FigConfOmegN5} the conformal constraint \eqref{confConstDE} as a function of $\rho$ and $\alpha$ is shown. It is evident that something different is happening and, moreover, the opposite conclusion should be extracted, this is that the region where PMS is realized is the worst case according to conformal symmetry.
To understand this we notice that ignoring all about conformal symmetry, the behavior is as for any other value of $N$, with one exception. This is, whether we look at the usual $\omega(\alpha)$ curve shown in Fig.~\ref{FigOmvsAlphaN5} or at the fixed point solutions for the functions $U_1$, $Z_1$ and $Y_1$, obtained from the dilatation Ward identity, as a function of $\rho$ and $\alpha$ (see, for example, the function $Z_1$ shown in Fig.~\ref{FigZ1N5}), all functions are regular. However, the function $\Upsilon$ presents a singular behavior, as shown in Fig.~\ref{FigUpsN5} where $\Upsilon'(\rho)$, as a function of the value of the $N$ parameter of the $O(N)$ symmetry, is represented for fixed values of $\rho$ and $\alpha$. It can be readily seen that a non-analyticity appears for a non-integer value of $N$, close to 4.5 for the chosen parameter $\alpha$.

It so happens that the dilatation equation which fixes $\Upsilon$ has the following structure:
\begin{equation}
	[2-D_\mathcal{O}+2D_\varphi+2D_\varphi\rho\partial_\rho-\mathbb{L}_1]\Upsilon'(\rho)=\mathbb{L}_0,
\end{equation}
where the functions $\mathbb{L}_0$ and $\mathbb{L}_1$ are the loops contributions which are functions of $\rho$ and depend on the fixed point solutions of $U_0^*$, $U_1^*$, $Z_0^*$, $Z_1^*$, $Y_0^*$ and $Y_1^*$. However, since all these functions (and also $D_\mathcal{O}$) are independent of $\Upsilon$, one is able to find $\Upsilon$ by inverting the operator:
\begin{equation}\label{UpsilonOp}
\mathbb{D}_\Upsilon\equiv2-D_\mathcal{O}+2D_\varphi+2D_\varphi\rho\partial_\rho-\mathbb{L}_1.
\end{equation}

\begin{figure}[h!]
	\centering
	\includegraphics[width=\columnwidth]{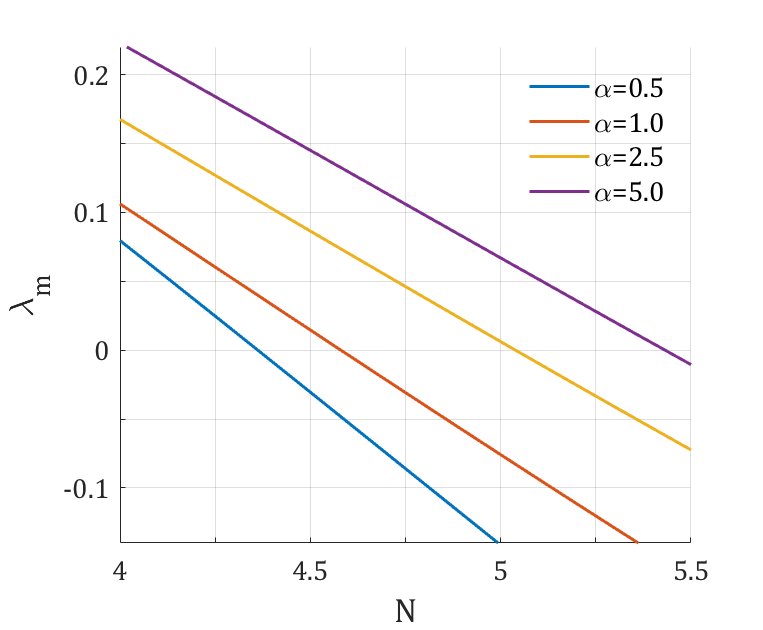}
	\caption{Smallest modulus eigenvalue of the operator defined in equation \eqref{confConstDE} when the least irrelevant perturbation is considered, as $N$ is increased from 4 to 5. The curves for various values of $\alpha$ were obtained employing the exponential regulator given in \eqref{regprofiles}. For every value of $\alpha$ considered in the range $[0.5,5]$, this eigenvalue presents a zero crossing in which the operator \eqref{confConstDE} becomes non-invertible, thus leading to an ill-defined $\Upsilon$ function, which in turn prevents PMC from being applied, as explained in the text.}
	\label{FigZeroEigN5}
\end{figure}

However, when analyzing the operator $\mathbb{D}_\Upsilon$ one finds that it has a zero eigenvalue and, consequently, it provokes a singular behavior in $\Upsilon$, see Fig.~\ref{FigZeroEigN5}, for a value of $N\approx 4.4$ which depends weakly on $\alpha$ as can be seen in Fig.~\ref{FigZeroEigN5}. Now, this singular behavior, by its own, does not affect the extraction of scaling dimensions from dilatation itself. This is because $\Upsilon$ does not feedback into the modified dilation equations that fix the other functions alongside with the scaling dimension $D_\mathcal{O}$. However, the function $\Upsilon$ has a very important role in the conformal constraint, see equation \eqref{confConstDE}. In particular, it enters in the left hand side, \textit{non-loop} term, which eventually end up spoiling the whole analysis of the conformal constraint. It is worth mentioning that this phenomenon affects also the values of $N=6$ and $N=7$ but to a lesser extent (in particular, there is a PMC value for $\omega$ for $N=7$) and, somehow, this does not seem to affect the case $N=4$. This is clearly an artifact of the DE approximation.

How to overcome this issue remains a question to be investigated. In particular, it must be studied how the effect of this problem evolves as one considers higher orders of the DE. A first possibility is that this \textit{zero} eigenvalue of the operator in equation \eqref{UpsilonOp} disappears. A second possibility may be that the degree of affectation of the singularity in $\Upsilon$ becomes narrower and does not affect integer values of $N$. One final hypothesis that we can think of is that this singularity moves to non-positive values of $N$ where the realization of conformal symmetry is not known to hold and, moreover, remaining as an indicator that, indeed, this is the case.

\begin{table}[!ht]
    \centering
    \begin{tabular}{l l l l}
    \toprule[1pt]\midrule[0.3pt]
          & $\nu$ & $\eta$  & $\omega$  \\
    \midrule     
    $\mathcal{O}\left(\partial^2\right)_{\text{fPMC}}$   & 0.782(9)   & & \\
    $\mathcal{O}\left(\partial^2\right)_{\text{fPMS}}$ & 0.7823(79)  & 0.0361(52) & 0.716(26)\\
    \midrule 
    $\mathcal{O}\left(\partial^2\right)_{\text{sPMS}}$   &  0.782(8) & 0.0364(52) & 0.724(34)\\
    $\mathcal{O}\left(\partial^4\right)_{\text{sPMS}}$   &  0.7797(9) & 0.0338(11) & 0.760(18)\\
    \midrule
    MC & 0.728(18) & & \\
    Large N  &  0.71(7) & 0.031(15) & 0.51(6)\\
    \midrule[0.3pt]\bottomrule[1pt]
    \end{tabular}
    \caption{Results obtained, with their corresponding error bars, for the $O(5)$ model. We report the predicted exponents through both the PMC and PMS criteria with the full version of the DE at order $\mathcal{O}\left(\partial^2\right)$for exponent $\nu$, and the PMS results for $\eta$ and $\omega$ -- in the text it is explained why there is not PMC prediction for $\omega$. We also include the results obtained with the strict version from \cite{DePolsi2020} and the results from MC calculations for $\nu$ \cite{Hu_2001} and the $1/N$ expansion to $(1/N^3)$ order.}
    \label{tab:N=5}
\end{table}

\subsection{Non-unitary models: the analytical extension to $N=-2$ and $N=0$}

As a final comment, we present our results for two interesting cases which constitute an analytical continuation of the $O(N)$ model equations: $N=0$ and $N=-2$. The results obtained in this work for these two non-unitary models are presented in Tables~\ref{tab:N=0} and \ref{tab:N=-2}. As was the case for all the preceding values of $N$ studied -- with the exemption of $N=5$ for $\omega$ -- we observe an excellent agreement between our PMS and PMC results, with central values differing by no more than $3\%$ and identical error bars. When compared with the strict results, however, we obtain slightly more precise results for the non-trivial exponent $\omega$ -- our error bars are slightly below one half of the corresponding strict version one's. As for the central values, once again both implementations of the DE are compatible well below the expected precision of the method at the present order. As a final remark, we would like to highlight that our results are compatible with discrepancies with the exact result for $\nu$ and the most precise perturbative result for $\omega$.

Our results pose a much interesting question. While for natural values of $N$ there exist clear mappings from the Euclidean classical $O(N)$-scalar models to unitary theories in Minkowskian spacetime, this is not the case for the $O(0)$ and $O(-2)$ models. For the former, as explained previously, there are very good reasons to expect the emergence of conformal symmetry in the critical regime, and this, as matter of fact, has indeed been proven for a few cases. Nevertheless, for the latter, even whether or not satisfying conformal symmetry has a physical interpretation is an open question, let alone understand how could it emerge in the critical regime. In view of this, having reasonable results coming from imposing the minimal breaking of the (analytical extension to non natural values of $N$ of the) SCT Ward identity suggests that conformal symmetry, in some yet-to-be understood sense, does take place in the critical regime and can we used to extract information about these non-unitary models. While for $\nu$ the $\alpha$ dependence is almost negligible, that it not the case for the $\omega$ exponent, and imposing the PMC for this exponent leads to essentially the same result than PMS -- in excellent agreement with the 6-loop calculation.

\begin{table}[!ht]
    \centering
    \begin{tabular}{l l l l}
    \toprule[1pt]\midrule[0.3pt]
          & $\nu$   & $\eta$ & $\omega$  \\
    \midrule     
    $\mathcal{O}\left(\partial^2\right)_{\text{fPMC}}$   & 0.5879(11) &   & 0.949(76)\\
    $\mathcal{O}\left(\partial^2\right)_{\text{fPMS}}$ & 0.5879(11) & 0.0330(47) & 0.946(76)\\
    \midrule 
    $\mathcal{O}\left(\partial^2\right)_{\text{sPMS}}$   &  0.5879(13) & 0.0326(47) & 1.00(19)\\
    $\mathcal{O}\left(\partial^4\right)_{\text{sPMS}}$   &  0.5876(2) & 0.0312(9) & 0.901(24)\\
    \midrule
    MC & 0.58759700(40) & 0.0310434(30) & 0.899(14) \\
    \midrule[0.3pt]\bottomrule[1pt]
    \end{tabular}
    \caption{Results obtained, with their corresponding error bars, for the $O(0)$ model. We report the predicted exponents through both the PMC and PMS criteria with the full version of the DE at order $\mathcal{O}\left(\partial^2\right)$. We also include the results obtained with the strict version from \cite{DePolsi2020} as well as the most precise results in the literature, coming from MC calculations \cite{Clisby16,Clisby_2017}.}
    \label{tab:N=0}
\end{table}

\begin{table}[!ht]
    \centering
    \begin{tabular}{l l l l}
    \toprule[1pt]\midrule[0.3pt]
    & $\nu$  & $\eta$ & $\omega$  \\
    \midrule     
     $\mathcal{O}\left(\partial^2\right)_{\text{fPMC}}$   & 0.5000(11)  &  & 0.818(76)\\
     $\mathcal{O}\left(\partial^2\right)_{\text{fPMS}}$ &0.5000(11)  & 0.0000(47) & 0.820(76)\\
    \midrule 
    $\mathcal{O}\left(\partial^2\right)_{\text{sPMS}}$   &   0.5000(12) & 0.0000(47) & 0.84(19)\\
    $\mathcal{O}\left(\partial^4\right)_{\text{sPMS}}$   &   0.5001(1) & 0.0004(9) & 0.838(24)\\
    \midrule
    exact/6-loop & 1/2 & 0 &  0.83(1)\\
    \midrule[0.3pt]\bottomrule[1pt]
  \end{tabular}
    \caption{Results obtained, with their corresponding error bars, for the $O(-2)$ model. We report the predicted exponents through both the PMC and PMS criteria with the full version of the DE at order $\mathcal{O}\left(\partial^2\right)$. We also include the results obtained with the strict version from \cite{DePolsi2020}. As a comparison benchmark, we include the exact values for $\nu$ and $\eta$, and a perturbative estimation for $\omega$ \cite{Wiese:2019xmu,Wiese:2018dow}.}
    \label{tab:N=-2}
\end{table}

\section{Conclusions}\label{secConcl}

In this work, we considered the modified Ward identities associated with conformal and dilatation symmetries in the context of the FRG, for the first time for the scalar $O(N)$ models extending the work done previously for the Ising universality class. We considered a source for a primary composite operator and studied the equations obtained at order $O(\partial^2)$ of the DE. As was the case for the Ising model, conformal invariance yields new information (with respect to scale invariance) already at this order of the approximation scheme. The equations arising when considering the Ward identities associated with conformal invariance are not fulfilled when evaluated at the fixed point, as opposed to what it is expected in the exact theory. This implies a constraint for each (primary) operator. We used the violation of these constraints, for the relevant and first irrelevant $O(N)$ invariant operator, in order to fix non-physical scheme parameters invoking the principle of maximal conformality introduced in previous works.

We have shown that for generic values of $N$ the PMC criterion is equivalent to using the PMS. Moreover, the results that these criteria yield are in agreement within error bars with each other and with others method reported in the literature. This also applies to the non-unitary cases such as $N=0$ and $N=-2$, where the realization of conformal symmetry remains an open question. An exception to this generic behavior occurs for values of $N\simeq 5$. As discussed, it so happens that the derivative expansion affects the conformal constraint, but not the PMS criterion for computing the $\omega$ critical exponent. This points to a failure of compatibility of the DE and conformal symmetry which could be resolved with higher orders of the approximation scheme, although how this may be resolved remains an open question to be tackled in the future.

This study showed that even though the PMS does not have such a solid first-principles argument as the PMC for critical properties, it is nevertheless equivalent to it. Moreover, given the fact that it is simpler and more robust, it makes it preferable in a general scenario, at least at this order of the DE approximation scheme. Finally, PMS can be applied to non-critical or non-universal properties at odds with the PMC.

One final byproduct of this work is that the two different implementations of the DE, namely the strict and the full version, agree within error bars, at least around the fixed point in $d=3$. This conclusion, allows for an implementation of the DE presenting equations which are much more compact which is a great simplification when considering higher orders of the DE where the size and complexity of the equations grow enormously.

\acknowledgments

We are very grateful to Matthieu Tissier for valuable comments on the manuscript.
This work received the support of the French-Uruguayan Institute of Physics project (IFU$\Phi$) and from the grant of number FCE-1-2021-1-166479 of the Agencia Nacional de Investigaci\'on e Innovaci\'on (Uruguay) and from the grant of number FVF/2021/160 from the Direcci\'on Nacional de Innovaci\'on, Ciencia y Tecnolog\'ia (Uruguay).

\appendix 

\section{Raw data obtained with the derivative expansion at order $\mathcal{O}(\partial^2)$}

In this Appendix we present the data as extracted directly from regulators and implementing the PMS and PMC criteria, but without any further processing.
\begin{table}[!ht]
    \vspace{0pt}
    \centering
    \begin{tabular}{l l l l l}
    \toprule[1pt]\midrule[0.3pt]
    & Regulator      & \multicolumn{1}{c}{$\nu$} & \multicolumn{1}{c}{$\eta$} & \multicolumn{1}{c}{$\omega$}  \\
    \midrule     
    PMC & $E$   & 0.62783  & & 0.84889 \\
        & $W$   & 0.62816  & & 0.84820 \\
    \midrule 
    PMS & $E$   & 0.62796   & 0.04477 & 0.84828 \\
        & $W$   & 0.62822   & 0.04426 & 0.84800 \\
    \midrule[0.3pt]\bottomrule[1pt]
    \end{tabular}
    \caption{Raw data  for the $O(1)$ model. We present PMS results for $\nu$, $\eta$ and $\omega$, and PMC values for $\nu$ and $\omega$, corresponding to the exponential and Wetterich regulators.}
    \label{tab:N=1raw}
\end{table}

\begin{table}[!ht]
    \vspace{0pt}
    \centering
    \begin{tabular}{l l l l l}
    \toprule[1pt]\midrule[0.3pt]
    & Regulator      & \multicolumn{1}{c}{$\nu$} & \multicolumn{1}{c}{$\eta$} & \multicolumn{1}{c}{$\omega$}\\
    \midrule     
    PMC & $E$   & 0.66675	& & 0.78480\\
        & $W$   & 0.66741	& & 0.78651\\
    \midrule 
    PMS & $E$   & 0.66695   & 0.04723 & 0.78597\\
        & $W$   & 0.66675   & 0.04667 & 0.78712\\
    \midrule[0.3pt]\bottomrule[1pt]
    \end{tabular}
    \caption{Raw data for the $O(2)$ model. We present PMS results for $\nu$, $\eta$ and $\omega$, and PMC values for $\nu$ and $\omega$, corresponding to the exponential $E$ and Wetterich $W$ regulators.}
    \label{tab:N=2raw}
\end{table}

\begin{table}[!ht]
    \vspace{0pt}
    \centering
    \begin{tabular}{l l l l l}
    \toprule[1pt]\midrule[0.3pt]
    & Regulator      & \multicolumn{1}{c}{$\nu$} & \multicolumn{1}{c}{$\eta$} & \multicolumn{1}{c}{$\omega$}\\
    \midrule     
    PMC & $E$   & 0.70452	& & 0.74290\\
        & $W$   & 0.70541   & & 0.74630\\
    \midrule 
    PMS & $E$   & 0.70473   & 0.04688 & 0.74463\\
        & $W$   & 0.70552   & 0.04632 & 0.74716\\
    \midrule[0.3pt]\bottomrule[1pt]
    \end{tabular}
    \caption{Raw data for the $O(3)$ model. We present PMS results for $\nu$, $\eta$ and $\omega$, and PMC values for $\nu$ and $\omega$, corresponding to the exponential $E$ and Wetterich $W$ regulators.}
    \label{tab:N=3raw}
\end{table}

\begin{table}[!ht]
    \vspace{0pt}
    \centering
    \begin{tabular}{l l l l l}
    \toprule[1pt]\midrule[0.3pt]
    & Regulator      & \multicolumn{1}{c}{$\nu$} & \multicolumn{1}{c}{$\eta$} & \multicolumn{1}{c}{$\omega$}\\
    \midrule     
    PMC & $E$   & 0.74037	& & 0.72054\\
        & $W$   & 0.74133	& & 0.72456\\
    \midrule 
    PMS & $E$   & 0.74056   & 0.04480 & 0.72145\\
        & $W$   & 0.74143   & 0.04426 & 0.72511\\
    \midrule[0.3pt]\bottomrule[1pt]
    \end{tabular}
    \caption{Raw data for the $O(4)$ model. We present PMS results for $\nu$, $\eta$ and $\omega$, and PMC values for $\nu$ and $\omega$, corresponding to the exponential $E$ and Wetterich $W$ regulators.}
    \label{tab:N=4raw}
\end{table}

\begin{table}[!ht]
    \vspace{0pt}
    \centering
    \begin{tabular}{l l l l l}
    \toprule[1pt]\midrule[0.3pt]
    & Regulator      & \multicolumn{1}{c}{$\nu$} & \multicolumn{1}{c}{$\eta$} & \multicolumn{1}{c}{$\omega$}\\
    \midrule     
    PMC & $E$   & 0.77322	& & \\
        & $W$   & 0.77410   & & \\
    \midrule 
    PMS & $E$   & 0.77338   & 0.04182 & 0.71399\\
        & $W$   & 0.77419   & 0.04131 & 0.71831\\
    \midrule[0.3pt]\bottomrule[1pt]
    \end{tabular}
    \caption{Raw data for the $O(5)$ model. We present PMS results for $\nu$, $\eta$ and $\omega$, and PMC values for $\nu$ and $\omega$, corresponding to the exponential $E$ and Wetterich $W$ regulators. Recall that as discussed in the main body, there is no prediction for $\omega$ from the PMC criterion for this value of $N$.}
    \label{tab:N=5raw}
\end{table}

\begin{table}[!ht]
    \vspace{0pt}
    \centering
    \begin{tabular}{l l l l l}
    \toprule[1pt]\midrule[0.3pt]
    & Regulator      & \multicolumn{1}{c}{$\nu$} & \multicolumn{1}{c}{$\eta$} & \multicolumn{1}{c}{$\omega$}\\
    \midrule     
    PMC & $E$   & 0.87877	&& 0.78058\\
        & $W$   & 0.87888	&& 0.78368\\
    \midrule 
    PMS & $E$   & 0.87906   & 0.02724 & 0.78110\\
        & $W$   & 0.87920   & 0.02697 & 0.78406\\
    \midrule[0.3pt]\bottomrule[1pt]
    \end{tabular}
    \caption{Raw data for the $O(10)$ model. We present PMS results for $\nu$, $\eta$ and $\omega$, and PMC values for $\nu$ and $\omega$, corresponding to the exponential $E$ and Wetterich $W$ regulators.}
    \label{tab:N=10raw}
\end{table}

\begin{table}[!ht]
    \vspace{0pt}
    \centering
    \begin{tabular}{l l l l l}
    \toprule[1pt]\midrule[0.3pt]
    & Regulator      & \multicolumn{1}{c}{$\nu$} & \multicolumn{1}{c}{$\eta$} & \multicolumn{1}{c}{$\omega$}\\
    \midrule     
    PMC & $E$   & 0.94300	&& 0.88541\\
        & $W$   & 0.94287   && 0.88624\\
    \midrule 
    PMS & $E$   & 0.94289   & 0.01456 & 0.88549\\
        & $W$   & 0.94278   & 0.01445 & 0.88625\\
    \midrule[0.3pt]\bottomrule[1pt]
    \end{tabular}
    \caption{Raw data for the $O(20)$ model. We present PMS results for $\nu$, $\eta$ and $\omega$, and PMC values for $\nu$ and $\omega$, corresponding to the exponential $E$ and Wetterich $W$ regulators.}
    \label{tab:N=20raw}
\end{table}

\begin{table}[!ht]
    \vspace{0pt}
    \centering
    \begin{tabular}{l l l l l}
    \toprule[1pt]\midrule[0.3pt]
    & Regulator      & \multicolumn{1}{c}{$\nu$} & \multicolumn{1}{c}{$\eta$} & \multicolumn{1}{c}{$\omega$}\\
    \midrule     
    PMC & $E$   & 0.98941	&& 0.97704\\
        & $W$   & 0.98937	&& 0.97723\\
    \midrule 
    PMS & $E$   & 0.98940   & 0.00297 & 0.97818\\
        & $W$   & 0.98937   & 0.00296 & 0.97821\\
    \midrule[0.3pt]\bottomrule[1pt]
    \end{tabular}
    \caption{Raw data for the $O(100)$ model. We present PMS results for $\nu$, $\eta$ and $\omega$, and PMC values for $\nu$ and $\omega$, corresponding to the exponential $E$ and Wetterich $W$ regulators.}
    \label{tab:N=100raw}
\end{table}

\begin{table}[!ht]
    \vspace{0pt}
    \centering
    \begin{tabular}{l l l l l}
    \toprule[1pt]\midrule[0.3pt]
    & Regulator      & \multicolumn{1}{c}{$\nu$} & \multicolumn{1}{c}{$\eta$} & \multicolumn{1}{c}{$\omega$}\\
    \midrule     
    PMC & $E$   & 0.58790	&& 0.95063\\
        & $W$   & 0.58792   && 0.94640\\
    \midrule 
    PMS & $E$   & 0.58789   & 0.03810 & 0.94743\\
        & $W$   & 0.58791   & 0.03768 & 0.94531\\
    \midrule[0.3pt]\bottomrule[1pt]
    \end{tabular}
    \caption{Raw data for the $O(0)$ model. We present PMS results for $\nu$, $\eta$ and $\omega$, and PMC values for $\nu$ and $\omega$, corresponding to the exponential $E$ and Wetterich $W$ regulators.}
    \label{tab:N=0raw}
\end{table}

\begin{table}[!ht]
    \vspace{0pt}
    \centering
    \begin{tabular}{l l l l l}
    \toprule[1pt]\midrule[0.3pt]
    & Regulator      & \multicolumn{1}{c}{$\nu$} & \multicolumn{1}{c}{$\eta$} & \multicolumn{1}{c}{$\omega$}\\
    \midrule     
    PMC & $E$   & 0.5+6.3$\times10^{-6}$	&& 0.81898\\
        & $W$   & 0.5+2.1$\times10^{-6}$	&& 0.81640\\
    \midrule 
    PMS & $E$   & 0.5+6.3$\times10^{-6}$   & $1.067\times 10^{-5}$ & 0.81833\\
        & $W$   & 0.5+2.1$\times10^{-6}$   &  & 0.82087\\
    \midrule[0.3pt]\bottomrule[1pt]
    \end{tabular}
    \caption{Raw data for the $O(-2)$ model. We present PMS results for $\nu$, $\eta$ and $\omega$, and PMC values for $\nu$ and $\omega$, corresponding to the exponential $E$ and Wetterich $W$ regulators.}
    \label{tab:N=-2raw}
\end{table}

\section{Numerical details}\label{Ap:numerics}

After deducing the expressions that were employed in this study through a symbolical routine, the numerical procedure consists in two stages. The first phase consists in solving the FRG flow equations in order to obtain the fixed point solution that governs the critical regime of the system. The second and final stage is that of performing a stability analysis around the fixed point or, equivalently, solving the equations for the eigenperturbations and, finally, to evaluate at this fixed point the conformal constraints for these eigenperturbations.

\subsection{Solving for the fixed point}

The FRG equations obtained for the different functions of the DE \textbf{ansatz}, in the presence of a generic regulator, are partial integro-differential equations. As such, we employ one the most ubiquitous methods to address differential equations: finite differences. To this end, we discretize the $\rho$ field dependence in a uniform grid, with varying number of points and maximum value, according to the value of $N$ considered -- with step $d\rho$ approximately constant. We implemented a 5-point centered derivative in the grid for the $\rho$ partial derivatives. We modified our definition towards the edges to maintain the number of points considered while remaining as centered as possible. For the loop integrals over $q$ momentum, we employed adaptive Gauss-Kronrod quadrature integration. We this setup, we started looking for a first estimation of the fixed point solution for $U_0$, $Z_0$ and $Y_0$, by a bi-partition method, starting from a microscopic initial condition and fine-tuning a temperature-like parameter until reaching a starting point whose flow remains constant a considerable `renormalization time', large enough so that reasonable proximity to the FRG fixed point can be assumed. An alternative procedure, which we did not employ in general (save for the $N=5$ case), consists on starting to work in a spatial dimension close to the upper critical one, finding the solution through a root-finding algorithm and then proceeding to gradually reduce the dimension towards the final value $d=3$. Both these procedures provide us with initial estimations for the fixed point-only dependent functions $U_0$, $Z_0$ and $Y_0$. 

The bi-partition method employed in this work, although conceptually clear, has a downside when precision is of the order. As the cumulative error grows with the flow from the microscopic scale $k=\Lambda$ up to the Ginzburg scale $k_G$, high levels of accuracy cannot be reached through this method alone. This means that a root-finding procedure to look for zeros of the flow equations must be implemented on top of the initial algorithm, which purpose then becomes to provide an educated initial estimate for this second step. For this work, the root-finding algorithm implemented consists on a discretized Newton-Raphson (also known as secant) method.

Since we are working at linear order in $K$, the equations fixing $U_1$, $Z_1$ and $Y_1$ (and also $D_O$) exhibit a linear dependence in themselves and consequently can be fixed up to a normalization constant (as is the case for an eigenvector). This normalization constant allows for the fixing of

\subsection{Linear analysis at the fixed point}

Once we are at the fixed point, up to a certain precision threshold, we need to solve the eigenvalue problem to find the eigenperturbations (which correspond to the functions $U_1$. $Z_1$, $Y_1$ and $\Upsilon$). We do this by resorting again to a Newton-Raphson procedure. For this purpose, a simple linear stability analysis of the fixed point serves as a very good starting point for the search of eigenvectors and eigenvalues. Once a desired threshold of convergence is achieved for the method, we conclude the numerical study by evaluating the conformal constraint with the found eigenvalue ($D_{\mathcal{O}}$), eigenvector 
(functions $\big\{U_1(\rho),Z_1(\rho),Y_1(\rho)\big\},\Upsilon(\rho)$) 
and the Wilson-Fisher fixed point solution (given by functions $\big\{U_0(\rho),Z_0(\rho),Y_0(\rho)\big\}$).

\bibliographystyle{apsrev4-2}
\bibliography{articleConformalComposite}

\begin{thebibliography}{63}%
\makeatletter
\providecommand \@ifxundefined [1]{%
 \@ifx{#1\undefined}
}%
\providecommand \@ifnum [1]{%
 \ifnum #1\expandafter \@firstoftwo
 \else \expandafter \@secondoftwo
 \fi
}%
\providecommand \@ifx [1]{%
 \ifx #1\expandafter \@firstoftwo
 \else \expandafter \@secondoftwo
 \fi
}%
\providecommand \natexlab [1]{#1}%
\providecommand \enquote  [1]{``#1''}%
\providecommand \bibnamefont  [1]{#1}%
\providecommand \bibfnamefont [1]{#1}%
\providecommand \citenamefont [1]{#1}%
\providecommand \href@noop [0]{\@secondoftwo}%
\providecommand \href [0]{\begingroup \@sanitize@url \@href}%
\providecommand \@href[1]{\@@startlink{#1}\@@href}%
\providecommand \@@href[1]{\endgroup#1\@@endlink}%
\providecommand \@sanitize@url [0]{\catcode `\\12\catcode `\$12\catcode
  `\&12\catcode `\#12\catcode `\^12\catcode `\_12\catcode `\%12\relax}%
\providecommand \@@startlink[1]{}%
\providecommand \@@endlink[0]{}%
\providecommand \url  [0]{\begingroup\@sanitize@url \@url }%
\providecommand \@url [1]{\endgroup\@href {#1}{\urlprefix }}%
\providecommand \urlprefix  [0]{URL }%
\providecommand \Eprint [0]{\href }%
\providecommand \doibase [0]{https://doi.org/}%
\providecommand \selectlanguage [0]{\@gobble}%
\providecommand \bibinfo  [0]{\@secondoftwo}%
\providecommand \bibfield  [0]{\@secondoftwo}%
\providecommand \translation [1]{[#1]}%
\providecommand \BibitemOpen [0]{}%
\providecommand \bibitemStop [0]{}%
\providecommand \bibitemNoStop [0]{.\EOS\space}%
\providecommand \EOS [0]{\spacefactor3000\relax}%
\providecommand \BibitemShut  [1]{\csname bibitem#1\endcsname}%
\let\auto@bib@innerbib\@empty
\bibitem [{\citenamefont {Wilson}\ and\ \citenamefont
  {Fisher}(1972)}]{Wilson:1971dc}%
  \BibitemOpen
  \bibfield  {author} {\bibinfo {author} {\bibfnamefont {K.~G.}\ \bibnamefont
  {Wilson}}\ and\ \bibinfo {author} {\bibfnamefont {M.~E.}\ \bibnamefont
  {Fisher}},\ }\href {https://doi.org/10.1103/PhysRevLett.28.240} {\bibfield
  {journal} {\bibinfo  {journal} {Phys. Rev. Lett.}\ }\textbf {\bibinfo
  {volume} {28}},\ \bibinfo {pages} {240} (\bibinfo {year} {1972})}\BibitemShut
  {NoStop}%
\bibitem [{\citenamefont {Wilson}\ and\ \citenamefont
  {Kogut}(1974)}]{Wilson:1973jj}%
  \BibitemOpen
  \bibfield  {author} {\bibinfo {author} {\bibfnamefont {K.~G.}\ \bibnamefont
  {Wilson}}\ and\ \bibinfo {author} {\bibfnamefont {J.~B.}\ \bibnamefont
  {Kogut}},\ }\href {https://doi.org/10.1016/0370-1573(74)90023-4} {\bibfield
  {journal} {\bibinfo  {journal} {Phys. Rept.}\ }\textbf {\bibinfo {volume}
  {12}},\ \bibinfo {pages} {75} (\bibinfo {year} {1974})}\BibitemShut {NoStop}%
\bibitem [{\citenamefont {Wetterich}(1993)}]{Wetterich:1992yh}%
  \BibitemOpen
  \bibfield  {author} {\bibinfo {author} {\bibfnamefont {C.}~\bibnamefont
  {Wetterich}},\ }\href
  {https://doi.org/https://doi.org/10.1016/0370-2693(93)90726-X} {\bibfield
  {journal} {\bibinfo  {journal} {Physics Letters B}\ }\textbf {\bibinfo
  {volume} {301}},\ \bibinfo {pages} {90 } (\bibinfo {year}
  {1993})}\BibitemShut {NoStop}%
\bibitem [{\citenamefont {Ellwanger}(1993)}]{Ellwanger:1993kk}%
  \BibitemOpen
  \bibfield  {author} {\bibinfo {author} {\bibfnamefont {U.}~\bibnamefont
  {Ellwanger}},\ }\href {https://doi.org/10.1007/BF01553022} {\bibfield
  {journal} {\bibinfo  {journal} {Z. Phys.}\ }\textbf {\bibinfo {volume}
  {C58}},\ \bibinfo {pages} {619} (\bibinfo {year} {1993})}\BibitemShut
  {NoStop}%
\bibitem [{\citenamefont {Morris}(1994)}]{Morris:1993qb}%
  \BibitemOpen
  \bibfield  {author} {\bibinfo {author} {\bibfnamefont {T.~R.}\ \bibnamefont
  {Morris}},\ }\href {https://doi.org/10.1142/S0217751X94000972} {\bibfield
  {journal} {\bibinfo  {journal} {Int. J. Mod. Phys.}\ }\textbf {\bibinfo
  {volume} {A9}},\ \bibinfo {pages} {2411} (\bibinfo {year} {1994})},\ \Eprint
  {https://arxiv.org/abs/hep-ph/9308265} {arXiv:hep-ph/9308265 [hep-ph]}
  \BibitemShut {NoStop}%
\bibitem [{\citenamefont {Delamotte}(2012)}]{Delamotte:2007pf}%
  \BibitemOpen
  \bibfield  {author} {\bibinfo {author} {\bibfnamefont {B.}~\bibnamefont
  {Delamotte}},\ }\href {https://doi.org/10.1007/978-3-642-27320-9_2}
  {\bibfield  {journal} {\bibinfo  {journal} {Lect. Notes Phys.}\ }\textbf
  {\bibinfo {volume} {852}},\ \bibinfo {pages} {49} (\bibinfo {year} {2012})},\
  \Eprint {https://arxiv.org/abs/cond-mat/0702365} {arXiv:cond-mat/0702365
  [cond-mat.stat-mech]} \BibitemShut {NoStop}%
\bibitem [{\citenamefont {Dupuis}\ \emph {et~al.}(2021)\citenamefont {Dupuis},
  \citenamefont {Canet}, \citenamefont {Eichhorn}, \citenamefont {Metzner},
  \citenamefont {Pawlowski}, \citenamefont {Tissier},\ and\ \citenamefont
  {Wschebor}}]{Dupuis:2020fhh}%
  \BibitemOpen
  \bibfield  {author} {\bibinfo {author} {\bibfnamefont {N.}~\bibnamefont
  {Dupuis}}, \bibinfo {author} {\bibfnamefont {L.}~\bibnamefont {Canet}},
  \bibinfo {author} {\bibfnamefont {A.}~\bibnamefont {Eichhorn}}, \bibinfo
  {author} {\bibfnamefont {W.}~\bibnamefont {Metzner}}, \bibinfo {author}
  {\bibfnamefont {J.~M.}\ \bibnamefont {Pawlowski}}, \bibinfo {author}
  {\bibfnamefont {M.}~\bibnamefont {Tissier}},\ and\ \bibinfo {author}
  {\bibfnamefont {N.}~\bibnamefont {Wschebor}},\ }\href
  {https://doi.org/10.1016/j.physrep.2021.01.001} {\bibfield  {journal}
  {\bibinfo  {journal} {Phys. Rept.}\ }\textbf {\bibinfo {volume} {910}},\
  \bibinfo {pages} {1} (\bibinfo {year} {2021})},\ \Eprint
  {https://arxiv.org/abs/2006.04853} {arXiv:2006.04853 [cond-mat.stat-mech]}
  \BibitemShut {NoStop}%
\bibitem [{\citenamefont {Polyakov}(1970)}]{Polyakov:1970xd}%
  \BibitemOpen
  \bibfield  {author} {\bibinfo {author} {\bibfnamefont {A.~M.}\ \bibnamefont
  {Polyakov}},\ }\href {https://ci.nii.ac.jp/naid/10006414995/en/} {\bibfield
  {journal} {\bibinfo  {journal} {JETP Lett.}\ }\textbf {\bibinfo {volume}
  {12}},\ \bibinfo {pages} {381} (\bibinfo {year} {1970})}\BibitemShut
  {NoStop}%
\bibitem [{\citenamefont {Migdal}(1971)}]{Migdal:1971xh}%
  \BibitemOpen
  \bibfield  {author} {\bibinfo {author} {\bibfnamefont {A.~A.}\ \bibnamefont
  {Migdal}},\ }\href {https://doi.org/10.1016/0370-2693(71)90583-1} {\bibfield
  {journal} {\bibinfo  {journal} {Phys. Lett.}\ }\textbf {\bibinfo {volume}
  {37B}},\ \bibinfo {pages} {98} (\bibinfo {year} {1971})}\BibitemShut
  {NoStop}%
\bibitem [{\citenamefont {Di~Francesco}\ \emph {et~al.}(1997)\citenamefont
  {Di~Francesco}, \citenamefont {Mathieu},\ and\ \citenamefont
  {Senechal}}]{DiFrancesco:1997nk}%
  \BibitemOpen
  \bibfield  {author} {\bibinfo {author} {\bibfnamefont {P.}~\bibnamefont
  {Di~Francesco}}, \bibinfo {author} {\bibfnamefont {P.}~\bibnamefont
  {Mathieu}},\ and\ \bibinfo {author} {\bibfnamefont {D.}~\bibnamefont
  {Senechal}},\ }\href {https://doi.org/10.1007/978-1-4612-2256-9} {\emph
  {\bibinfo {title} {{Conformal Field Theory}}}},\ Graduate Texts in
  Contemporary Physics\ (\bibinfo  {publisher} {Springer-Verlag},\ \bibinfo
  {address} {New York},\ \bibinfo {year} {1997})\BibitemShut {NoStop}%
\bibitem [{\citenamefont {Belavin}\ \emph {et~al.}(1984)\citenamefont
  {Belavin}, \citenamefont {Polyakov},\ and\ \citenamefont
  {Zamolodchikov}}]{Belavin:1984vu}%
  \BibitemOpen
  \bibfield  {author} {\bibinfo {author} {\bibfnamefont {A.~A.}\ \bibnamefont
  {Belavin}}, \bibinfo {author} {\bibfnamefont {A.~M.}\ \bibnamefont
  {Polyakov}},\ and\ \bibinfo {author} {\bibfnamefont {A.~B.}\ \bibnamefont
  {Zamolodchikov}},\ }\href {https://doi.org/10.1016/0550-3213(84)90052-X}
  {\bibfield  {journal} {\bibinfo  {journal} {Nucl. Phys.}\ }\textbf {\bibinfo
  {volume} {B241}},\ \bibinfo {pages} {333} (\bibinfo {year} {1984})},\
  \bibinfo {note} {[,605(1984)]}\BibitemShut {NoStop}%
\bibitem [{\citenamefont {Delamotte}\ \emph {et~al.}(2016)\citenamefont
  {Delamotte}, \citenamefont {Tissier},\ and\ \citenamefont
  {Wschebor}}]{delamotte2016scale}%
  \BibitemOpen
  \bibfield  {author} {\bibinfo {author} {\bibfnamefont {B.}~\bibnamefont
  {Delamotte}}, \bibinfo {author} {\bibfnamefont {M.}~\bibnamefont {Tissier}},\
  and\ \bibinfo {author} {\bibfnamefont {N.}~\bibnamefont {Wschebor}},\
  }\href@noop {} {\bibfield  {journal} {\bibinfo  {journal} {Physical Review
  E}\ }\textbf {\bibinfo {volume} {93}},\ \bibinfo {pages} {012144} (\bibinfo
  {year} {2016})}\BibitemShut {NoStop}%
\bibitem [{\citenamefont {De~Polsi}\ \emph {et~al.}(2019)\citenamefont
  {De~Polsi}, \citenamefont {Tissier},\ and\ \citenamefont
  {Wschebor}}]{DePolsi2019}%
  \BibitemOpen
  \bibfield  {author} {\bibinfo {author} {\bibfnamefont {G.}~\bibnamefont
  {De~Polsi}}, \bibinfo {author} {\bibfnamefont {M.}~\bibnamefont {Tissier}},\
  and\ \bibinfo {author} {\bibfnamefont {N.}~\bibnamefont {Wschebor}},\ }\href
  {https://doi.org/10.1007/s10955-019-02411-3} {\bibfield  {journal} {\bibinfo
  {journal} {Journal of Statistical Physics}\ }\textbf {\bibinfo {volume}
  {177}},\ \bibinfo {pages} {1089} (\bibinfo {year} {2019})}\BibitemShut
  {NoStop}%
\bibitem [{\citenamefont {Hasenbusch}(2019)}]{Hasenbusch:2019jkj}%
  \BibitemOpen
  \bibfield  {author} {\bibinfo {author} {\bibfnamefont {M.}~\bibnamefont
  {Hasenbusch}},\ }\href {https://doi.org/10.1103/PhysRevB.100.224517}
  {\bibfield  {journal} {\bibinfo  {journal} {Phys. Rev. B}\ }\textbf {\bibinfo
  {volume} {100}},\ \bibinfo {pages} {224517} (\bibinfo {year}
  {2019})}\BibitemShut {NoStop}%
\bibitem [{\citenamefont {{Hasenbusch}}(2020)}]{Hasenbusch2005}%
  \BibitemOpen
  \bibfield  {author} {\bibinfo {author} {\bibfnamefont {M.}~\bibnamefont
  {{Hasenbusch}}},\ }\href {https://doi.org/10.1103/PhysRevB.102.024406}
  {\bibfield  {journal} {\bibinfo  {journal} {\prb}\ }\textbf {\bibinfo
  {volume} {102}},\ \bibinfo {eid} {024406} (\bibinfo {year} {2020})},\ \Eprint
  {https://arxiv.org/abs/2005.04448} {arXiv:2005.04448 [cond-mat.stat-mech]}
  \BibitemShut {NoStop}%
\bibitem [{\citenamefont {Hasenbusch}(2021)}]{Hasenbusch:2021tei}%
  \BibitemOpen
  \bibfield  {author} {\bibinfo {author} {\bibfnamefont {M.}~\bibnamefont
  {Hasenbusch}},\ }\href {https://doi.org/10.1103/PhysRevB.104.014426}
  {\bibfield  {journal} {\bibinfo  {journal} {Phys. Rev. B}\ }\textbf {\bibinfo
  {volume} {104}},\ \bibinfo {pages} {014426} (\bibinfo {year} {2021})},\
  \Eprint {https://arxiv.org/abs/2105.09781} {arXiv:2105.09781
  [cond-mat.stat-mech]} \BibitemShut {NoStop}%
\bibitem [{\citenamefont {Schnetz}(2018)}]{Schnetz:2016fhy}%
  \BibitemOpen
  \bibfield  {author} {\bibinfo {author} {\bibfnamefont {O.}~\bibnamefont
  {Schnetz}},\ }\href {https://doi.org/10.1103/PhysRevD.97.085018} {\bibfield
  {journal} {\bibinfo  {journal} {Phys. Rev. D}\ }\textbf {\bibinfo {volume}
  {97}},\ \bibinfo {pages} {085018} (\bibinfo {year} {2018})},\ \Eprint
  {https://arxiv.org/abs/1606.08598} {arXiv:1606.08598 [hep-th]} \BibitemShut
  {NoStop}%
\bibitem [{\citenamefont {Kompaniets}\ and\ \citenamefont
  {Panzer}(2017)}]{Kompaniets:2017yct}%
  \BibitemOpen
  \bibfield  {author} {\bibinfo {author} {\bibfnamefont {M.~V.}\ \bibnamefont
  {Kompaniets}}\ and\ \bibinfo {author} {\bibfnamefont {E.}~\bibnamefont
  {Panzer}},\ }\href {https://doi.org/10.1103/PhysRevD.96.036016} {\bibfield
  {journal} {\bibinfo  {journal} {Phys. Rev.}\ }\textbf {\bibinfo {volume}
  {D96}},\ \bibinfo {pages} {036016} (\bibinfo {year} {2017})},\ \Eprint
  {https://arxiv.org/abs/1705.06483} {arXiv:1705.06483 [hep-th]} \BibitemShut
  {NoStop}%
\bibitem [{\citenamefont {Shalaby}(2020)}]{Shalaby:2020faz}%
  \BibitemOpen
  \bibfield  {author} {\bibinfo {author} {\bibfnamefont {A.~M.}\ \bibnamefont
  {Shalaby}},\ }\href {https://doi.org/10.1103/PhysRevD.102.105017} {\bibfield
  {journal} {\bibinfo  {journal} {Phys. Rev. D}\ }\textbf {\bibinfo {volume}
  {102}},\ \bibinfo {pages} {105017} (\bibinfo {year} {2020})},\ \Eprint
  {https://arxiv.org/abs/2010.13097} {arXiv:2010.13097 [hep-th]} \BibitemShut
  {NoStop}%
\bibitem [{\citenamefont {{Abhignan}}\ and\ \citenamefont
  {{Sankaranarayanan}}(2021)}]{Abhignan06}%
  \BibitemOpen
  \bibfield  {author} {\bibinfo {author} {\bibfnamefont {V.}~\bibnamefont
  {{Abhignan}}}\ and\ \bibinfo {author} {\bibfnamefont {R.}~\bibnamefont
  {{Sankaranarayanan}}},\ }\href {https://doi.org/10.1007/s10955-021-02719-z}
  {\bibfield  {journal} {\bibinfo  {journal} {Journal of Statistical Physics}\
  }\textbf {\bibinfo {volume} {183}},\ \bibinfo {eid} {4} (\bibinfo {year}
  {2021})},\ \Eprint {https://arxiv.org/abs/2006.12064} {arXiv:2006.12064
  [cond-mat.stat-mech]} \BibitemShut {NoStop}%
\bibitem [{\citenamefont {Shalaby}(2021)}]{Shalaby:2020xvv}%
  \BibitemOpen
  \bibfield  {author} {\bibinfo {author} {\bibfnamefont {A.~M.}\ \bibnamefont
  {Shalaby}},\ }\href {https://doi.org/10.1140/epjc/s10052-021-08884-5}
  {\bibfield  {journal} {\bibinfo  {journal} {Eur. Phys. J. C}\ }\textbf
  {\bibinfo {volume} {81}},\ \bibinfo {pages} {87} (\bibinfo {year} {2021})},\
  \Eprint {https://arxiv.org/abs/2005.12714} {arXiv:2005.12714 [hep-th]}
  \BibitemShut {NoStop}%
\bibitem [{\citenamefont {El-Showk}\ \emph {et~al.}(2014)\citenamefont
  {El-Showk}, \citenamefont {Paulos}, \citenamefont {Poland}, \citenamefont
  {Rychkov}, \citenamefont {Simmons-Duffin},\ and\ \citenamefont
  {Vichi}}]{El-Showk:2014dwa}%
  \BibitemOpen
  \bibfield  {author} {\bibinfo {author} {\bibfnamefont {S.}~\bibnamefont
  {El-Showk}}, \bibinfo {author} {\bibfnamefont {M.~F.}\ \bibnamefont
  {Paulos}}, \bibinfo {author} {\bibfnamefont {D.}~\bibnamefont {Poland}},
  \bibinfo {author} {\bibfnamefont {S.}~\bibnamefont {Rychkov}}, \bibinfo
  {author} {\bibfnamefont {D.}~\bibnamefont {Simmons-Duffin}},\ and\ \bibinfo
  {author} {\bibfnamefont {A.}~\bibnamefont {Vichi}},\ }\href
  {https://doi.org/10.1007/s10955-014-1042-7} {\bibfield  {journal} {\bibinfo
  {journal} {J. Stat. Phys.}\ }\textbf {\bibinfo {volume} {157}},\ \bibinfo
  {pages} {869} (\bibinfo {year} {2014})},\ \Eprint
  {https://arxiv.org/abs/1403.4545} {arXiv:1403.4545 [hep-th]} \BibitemShut
  {NoStop}%
\bibitem [{\citenamefont {El-Showk}\ \emph {et~al.}(2012)\citenamefont
  {El-Showk}, \citenamefont {Paulos}, \citenamefont {Poland}, \citenamefont
  {Rychkov}, \citenamefont {Simmons-Duffin},\ and\ \citenamefont
  {Vichi}}]{ElShowk:2012ht}%
  \BibitemOpen
  \bibfield  {author} {\bibinfo {author} {\bibfnamefont {S.}~\bibnamefont
  {El-Showk}}, \bibinfo {author} {\bibfnamefont {M.~F.}\ \bibnamefont
  {Paulos}}, \bibinfo {author} {\bibfnamefont {D.}~\bibnamefont {Poland}},
  \bibinfo {author} {\bibfnamefont {S.}~\bibnamefont {Rychkov}}, \bibinfo
  {author} {\bibfnamefont {D.}~\bibnamefont {Simmons-Duffin}},\ and\ \bibinfo
  {author} {\bibfnamefont {A.}~\bibnamefont {Vichi}},\ }\href
  {https://doi.org/10.1103/PhysRevD.86.025022} {\bibfield  {journal} {\bibinfo
  {journal} {Phys. Rev.}\ }\textbf {\bibinfo {volume} {D86}},\ \bibinfo {pages}
  {025022} (\bibinfo {year} {2012})},\ \Eprint
  {https://arxiv.org/abs/1203.6064} {arXiv:1203.6064 [hep-th]} \BibitemShut
  {NoStop}%
\bibitem [{\citenamefont {Kos}\ \emph {et~al.}(2014)\citenamefont {Kos},
  \citenamefont {Poland},\ and\ \citenamefont {Simmons-Duffin}}]{Kos:2014bka}%
  \BibitemOpen
  \bibfield  {author} {\bibinfo {author} {\bibfnamefont {F.}~\bibnamefont
  {Kos}}, \bibinfo {author} {\bibfnamefont {D.}~\bibnamefont {Poland}},\ and\
  \bibinfo {author} {\bibfnamefont {D.}~\bibnamefont {Simmons-Duffin}},\ }\href
  {https://doi.org/10.1007/JHEP11(2014)109} {\bibfield  {journal} {\bibinfo
  {journal} {JHEP}\ }\textbf {\bibinfo {volume} {11}},\ \bibinfo {pages}
  {109}},\ \Eprint {https://arxiv.org/abs/1406.4858} {arXiv:1406.4858 [hep-th]}
  \BibitemShut {NoStop}%
\bibitem [{\citenamefont {Chester}\ \emph {et~al.}(2019)\citenamefont
  {Chester}, \citenamefont {Landry}, \citenamefont {Liu}, \citenamefont
  {Poland}, \citenamefont {Simmons-Duffin}, \citenamefont {Su},\ and\
  \citenamefont {Vichi}}]{chester2019carving}%
  \BibitemOpen
  \bibfield  {author} {\bibinfo {author} {\bibfnamefont {S.~M.}\ \bibnamefont
  {Chester}}, \bibinfo {author} {\bibfnamefont {W.}~\bibnamefont {Landry}},
  \bibinfo {author} {\bibfnamefont {J.}~\bibnamefont {Liu}}, \bibinfo {author}
  {\bibfnamefont {D.}~\bibnamefont {Poland}}, \bibinfo {author} {\bibfnamefont
  {D.}~\bibnamefont {Simmons-Duffin}}, \bibinfo {author} {\bibfnamefont
  {N.}~\bibnamefont {Su}},\ and\ \bibinfo {author} {\bibfnamefont
  {A.}~\bibnamefont {Vichi}},\ }\href@noop {} {\bibinfo {title} {Carving out
  ope space and precise $o(2)$ model critical exponents}} (\bibinfo {year}
  {2019}),\ \Eprint {https://arxiv.org/abs/1912.03324} {arXiv:1912.03324
  [hep-th]} \BibitemShut {NoStop}%
\bibitem [{\citenamefont {Meneses}\ \emph {et~al.}(2019)\citenamefont
  {Meneses}, \citenamefont {Penedones}, \citenamefont {Rychkov}, \citenamefont
  {Viana Parente~Lopes},\ and\ \citenamefont
  {Yvernay}}]{meneses2019structural}%
  \BibitemOpen
  \bibfield  {author} {\bibinfo {author} {\bibfnamefont {S.}~\bibnamefont
  {Meneses}}, \bibinfo {author} {\bibfnamefont {J.}~\bibnamefont {Penedones}},
  \bibinfo {author} {\bibfnamefont {S.}~\bibnamefont {Rychkov}}, \bibinfo
  {author} {\bibfnamefont {J.}~\bibnamefont {Viana Parente~Lopes}},\ and\
  \bibinfo {author} {\bibfnamefont {P.}~\bibnamefont {Yvernay}},\ }\href@noop
  {} {\bibfield  {journal} {\bibinfo  {journal} {Journal of High Energy
  Physics}\ }\textbf {\bibinfo {volume} {2019}},\ \bibinfo {pages} {1}
  (\bibinfo {year} {2019})}\BibitemShut {NoStop}%
\bibitem [{\citenamefont {De~Polsi}\ \emph {et~al.}(2020)\citenamefont
  {De~Polsi}, \citenamefont {Balog}, \citenamefont {Tissier},\ and\
  \citenamefont {Wschebor}}]{DePolsi2020}%
  \BibitemOpen
  \bibfield  {author} {\bibinfo {author} {\bibfnamefont {G.}~\bibnamefont
  {De~Polsi}}, \bibinfo {author} {\bibfnamefont {I.}~\bibnamefont {Balog}},
  \bibinfo {author} {\bibfnamefont {M.}~\bibnamefont {Tissier}},\ and\ \bibinfo
  {author} {\bibfnamefont {N.}~\bibnamefont {Wschebor}},\ }\href
  {https://doi.org/10.1103/PhysRevE.101.042113} {\bibfield  {journal} {\bibinfo
   {journal} {Phys. Rev. E}\ }\textbf {\bibinfo {volume} {101}},\ \bibinfo
  {pages} {042113} (\bibinfo {year} {2020})}\BibitemShut {NoStop}%
\bibitem [{\citenamefont {De~Polsi}\ \emph {et~al.}(2021)\citenamefont
  {De~Polsi}, \citenamefont {Hern\'andez-Chifflet},\ and\ \citenamefont
  {Wschebor}}]{DePolsi2021}%
  \BibitemOpen
  \bibfield  {author} {\bibinfo {author} {\bibfnamefont {G.}~\bibnamefont
  {De~Polsi}}, \bibinfo {author} {\bibfnamefont {G.}~\bibnamefont
  {Hern\'andez-Chifflet}},\ and\ \bibinfo {author} {\bibfnamefont
  {N.}~\bibnamefont {Wschebor}},\ }\href
  {https://doi.org/10.1103/PhysRevE.104.064101} {\bibfield  {journal} {\bibinfo
   {journal} {Phys. Rev. E}\ }\textbf {\bibinfo {volume} {104}},\ \bibinfo
  {pages} {064101} (\bibinfo {year} {2021})}\BibitemShut {NoStop}%
\bibitem [{\citenamefont {P\'eli}(2021)}]{Peli:2020yiz}%
  \BibitemOpen
  \bibfield  {author} {\bibinfo {author} {\bibfnamefont {Z.}~\bibnamefont
  {P\'eli}},\ }\href {https://doi.org/10.1103/PhysRevE.103.032135} {\bibfield
  {journal} {\bibinfo  {journal} {Phys. Rev. E}\ }\textbf {\bibinfo {volume}
  {103}},\ \bibinfo {pages} {032135} (\bibinfo {year} {2021})},\ \Eprint
  {https://arxiv.org/abs/2010.04020} {arXiv:2010.04020 [hep-th]} \BibitemShut
  {NoStop}%
\bibitem [{\citenamefont {S\'anchez-Villalobos}\ \emph
  {et~al.}(2023)\citenamefont {S\'anchez-Villalobos}, \citenamefont
  {Delamotte},\ and\ \citenamefont {Wschebor}}]{Sanchez2023}%
  \BibitemOpen
  \bibfield  {author} {\bibinfo {author} {\bibfnamefont {C.~A.}\ \bibnamefont
  {S\'anchez-Villalobos}}, \bibinfo {author} {\bibfnamefont {B.}~\bibnamefont
  {Delamotte}},\ and\ \bibinfo {author} {\bibfnamefont {N.}~\bibnamefont
  {Wschebor}},\ }\href {https://doi.org/10.1103/PhysRevE.108.064120} {\bibfield
   {journal} {\bibinfo  {journal} {Phys. Rev. E}\ }\textbf {\bibinfo {volume}
  {108}},\ \bibinfo {pages} {064120} (\bibinfo {year} {2023})}\BibitemShut
  {NoStop}%
\bibitem [{\citenamefont {Johnson}\ \emph {et~al.}(2023)\citenamefont
  {Johnson}, \citenamefont {Rennecke},\ and\ \citenamefont
  {Skokov}}]{Rennecke2024}%
  \BibitemOpen
  \bibfield  {author} {\bibinfo {author} {\bibfnamefont {G.}~\bibnamefont
  {Johnson}}, \bibinfo {author} {\bibfnamefont {F.}~\bibnamefont {Rennecke}},\
  and\ \bibinfo {author} {\bibfnamefont {V.~V.}\ \bibnamefont {Skokov}},\
  }\href {https://doi.org/10.1103/PhysRevD.107.116013} {\bibfield  {journal}
  {\bibinfo  {journal} {Phys. Rev. D}\ }\textbf {\bibinfo {volume} {107}},\
  \bibinfo {pages} {116013} (\bibinfo {year} {2023})}\BibitemShut {NoStop}%
\bibitem [{\citenamefont {Sánchez-Villalobos}\ \emph
  {et~al.}(2024)\citenamefont {Sánchez-Villalobos}, \citenamefont
  {Delamotte},\ and\ \citenamefont {Wschebor}}]{Sanchez2024}%
  \BibitemOpen
  \bibfield  {author} {\bibinfo {author} {\bibfnamefont {C.~A.}\ \bibnamefont
  {Sánchez-Villalobos}}, \bibinfo {author} {\bibfnamefont {B.}~\bibnamefont
  {Delamotte}},\ and\ \bibinfo {author} {\bibfnamefont {N.}~\bibnamefont
  {Wschebor}},\ }\href {https://arxiv.org/abs/2411.02616} {\bibinfo {title}
  {$o(n)\times o(2)$ scalar models: including $\mathcal{O}(\partial^2)$
  corrections in the functional renormalization group analysis}} (\bibinfo
  {year} {2024}),\ \Eprint {https://arxiv.org/abs/2411.02616} {arXiv:2411.02616
  [cond-mat.stat-mech]} \BibitemShut {NoStop}%
\bibitem [{\citenamefont {Balog}\ \emph
  {et~al.}(2019{\natexlab{a}})\citenamefont {Balog}, \citenamefont
  {Chat{\'{e}}}, \citenamefont {Delamotte}, \citenamefont {Marohnic},
  \citenamefont {Wschebor}, \citenamefont {Marohni{\'{c}}},\ and\ \citenamefont
  {Wschebor}}]{Balog2019}%
  \BibitemOpen
  \bibfield  {author} {\bibinfo {author} {\bibfnamefont {I.}~\bibnamefont
  {Balog}}, \bibinfo {author} {\bibfnamefont {H.}~\bibnamefont {Chat{\'{e}}}},
  \bibinfo {author} {\bibfnamefont {B.}~\bibnamefont {Delamotte}}, \bibinfo
  {author} {\bibfnamefont {M.}~\bibnamefont {Marohnic}}, \bibinfo {author}
  {\bibfnamefont {N.}~\bibnamefont {Wschebor}}, \bibinfo {author}
  {\bibfnamefont {M.}~\bibnamefont {Marohni{\'{c}}}},\ and\ \bibinfo {author}
  {\bibfnamefont {N.}~\bibnamefont {Wschebor}},\ }\href
  {https://doi.org/10.1103/PhysRevLett.123.240604} {\bibfield  {journal}
  {\bibinfo  {journal} {Phys. Rev. Lett.}\ }\textbf {\bibinfo {volume} {123}},\
  \bibinfo {pages} {240604} (\bibinfo {year} {2019}{\natexlab{a}})},\ \Eprint
  {https://arxiv.org/abs/1907.01829} {arXiv:1907.01829 [cond-mat.stat-mech]}
  \BibitemShut {NoStop}%
\bibitem [{\citenamefont {Chlebicki}\ \emph {et~al.}(2022)\citenamefont
  {Chlebicki}, \citenamefont {S\'anchez-Villalobos}, \citenamefont
  {Jakubczyk},\ and\ \citenamefont {Wschebor}}]{Chlebicki:2022pxm}%
  \BibitemOpen
  \bibfield  {author} {\bibinfo {author} {\bibfnamefont {A.}~\bibnamefont
  {Chlebicki}}, \bibinfo {author} {\bibfnamefont {C.~A.}\ \bibnamefont
  {S\'anchez-Villalobos}}, \bibinfo {author} {\bibfnamefont {P.}~\bibnamefont
  {Jakubczyk}},\ and\ \bibinfo {author} {\bibfnamefont {N.}~\bibnamefont
  {Wschebor}},\ }\href {https://doi.org/10.1103/PhysRevE.106.064135} {\bibfield
   {journal} {\bibinfo  {journal} {Phys. Rev. E}\ }\textbf {\bibinfo {volume}
  {106}},\ \bibinfo {pages} {064135} (\bibinfo {year} {2022})},\ \Eprint
  {https://arxiv.org/abs/2204.02089} {arXiv:2204.02089 [cond-mat.stat-mech]}
  \BibitemShut {NoStop}%
\bibitem [{\citenamefont {S\'anchez-Villalobos}\ \emph
  {et~al.}(2024)\citenamefont {S\'anchez-Villalobos}, \citenamefont
  {Delamotte},\ and\ \citenamefont {Wschebor}}]{Sanchez-Villalobos:2024vmd}%
  \BibitemOpen
  \bibfield  {author} {\bibinfo {author} {\bibfnamefont {C.~A.}\ \bibnamefont
  {S\'anchez-Villalobos}}, \bibinfo {author} {\bibfnamefont {B.}~\bibnamefont
  {Delamotte}},\ and\ \bibinfo {author} {\bibfnamefont {N.}~\bibnamefont
  {Wschebor}},\ }\href@noop {} {\bibinfo {title} {{$O(N)\times O(2)$ scalar
  models: including $\mathcal{O}(\partial^2)$ corrections in the Functional
  Renormalization Group analysis}}} (\bibinfo {year} {2024}),\ \Eprint
  {https://arxiv.org/abs/2411.02616} {arXiv:2411.02616 [cond-mat.stat-mech]}
  \BibitemShut {NoStop}%
\bibitem [{\citenamefont {De~Polsi}\ and\ \citenamefont
  {Wschebor}(2022)}]{DePolsi2022}%
  \BibitemOpen
  \bibfield  {author} {\bibinfo {author} {\bibfnamefont {G.}~\bibnamefont
  {De~Polsi}}\ and\ \bibinfo {author} {\bibfnamefont {N.}~\bibnamefont
  {Wschebor}},\ }\href {https://doi.org/10.1103/PhysRevE.106.024111} {\bibfield
   {journal} {\bibinfo  {journal} {Phys. Rev. E}\ }\textbf {\bibinfo {volume}
  {106}},\ \bibinfo {pages} {024111} (\bibinfo {year} {2022})}\BibitemShut
  {NoStop}%
\bibitem [{\citenamefont {Balog}\ \emph {et~al.}(2020)\citenamefont {Balog},
  \citenamefont {De~Polsi}, \citenamefont {Tissier},\ and\ \citenamefont
  {Wschebor}}]{Balog2020}%
  \BibitemOpen
  \bibfield  {author} {\bibinfo {author} {\bibfnamefont {I.}~\bibnamefont
  {Balog}}, \bibinfo {author} {\bibfnamefont {G.}~\bibnamefont {De~Polsi}},
  \bibinfo {author} {\bibfnamefont {M.}~\bibnamefont {Tissier}},\ and\ \bibinfo
  {author} {\bibfnamefont {N.}~\bibnamefont {Wschebor}},\ }\href
  {https://doi.org/10.1103/PhysRevE.101.062146} {\bibfield  {journal} {\bibinfo
   {journal} {Phys. Rev. E}\ }\textbf {\bibinfo {volume} {101}},\ \bibinfo
  {pages} {062146} (\bibinfo {year} {2020})}\BibitemShut {NoStop}%
\bibitem [{\citenamefont {Delamotte}\ \emph {et~al.}(2024)\citenamefont
  {Delamotte}, \citenamefont {De~Polsi}, \citenamefont {Tissier},\ and\
  \citenamefont {Wschebor}}]{Delamotte2024}%
  \BibitemOpen
  \bibfield  {author} {\bibinfo {author} {\bibfnamefont {B.}~\bibnamefont
  {Delamotte}}, \bibinfo {author} {\bibfnamefont {G.}~\bibnamefont {De~Polsi}},
  \bibinfo {author} {\bibfnamefont {M.}~\bibnamefont {Tissier}},\ and\ \bibinfo
  {author} {\bibfnamefont {N.}~\bibnamefont {Wschebor}},\ }\href
  {https://doi.org/10.1103/PhysRevE.109.064152} {\bibfield  {journal} {\bibinfo
   {journal} {Phys. Rev. E}\ }\textbf {\bibinfo {volume} {109}},\ \bibinfo
  {pages} {064152} (\bibinfo {year} {2024})}\BibitemShut {NoStop}%
\bibitem [{\citenamefont {Polchinski}(1984)}]{Polchinski:1983gv}%
  \BibitemOpen
  \bibfield  {author} {\bibinfo {author} {\bibfnamefont {J.}~\bibnamefont
  {Polchinski}},\ }\href {https://doi.org/10.1016/0550-3213(84)90287-6}
  {\bibfield  {journal} {\bibinfo  {journal} {Nucl. Phys.}\ }\textbf {\bibinfo
  {volume} {B231}},\ \bibinfo {pages} {269} (\bibinfo {year}
  {1984})}\BibitemShut {NoStop}%
\bibitem [{\citenamefont {Berges}\ \emph
  {et~al.}(2002{\natexlab{a}})\citenamefont {Berges}, \citenamefont
  {Tetradis},\ and\ \citenamefont {Wetterich}}]{Berges:2000ew}%
  \BibitemOpen
  \bibfield  {author} {\bibinfo {author} {\bibfnamefont {J.}~\bibnamefont
  {Berges}}, \bibinfo {author} {\bibfnamefont {N.}~\bibnamefont {Tetradis}},\
  and\ \bibinfo {author} {\bibfnamefont {C.}~\bibnamefont {Wetterich}},\ }\href
  {https://doi.org/10.1016/S0370-1573(01)00098-9} {\bibfield  {journal}
  {\bibinfo  {journal} {Phys. Rept.}\ }\textbf {\bibinfo {volume} {363}},\
  \bibinfo {pages} {223} (\bibinfo {year} {2002}{\natexlab{a}})},\ \Eprint
  {https://arxiv.org/abs/hep-ph/0005122} {arXiv:hep-ph/0005122 [hep-ph]}
  \BibitemShut {NoStop}%
\bibitem [{\citenamefont {Pawlowski}(2007)}]{Pawlowski:2005xe}%
  \BibitemOpen
  \bibfield  {author} {\bibinfo {author} {\bibfnamefont {J.~M.}\ \bibnamefont
  {Pawlowski}},\ }\href {https://doi.org/10.1016/j.aop.2007.01.007} {\bibfield
  {journal} {\bibinfo  {journal} {Annals Phys.}\ }\textbf {\bibinfo {volume}
  {322}},\ \bibinfo {pages} {2831} (\bibinfo {year} {2007})},\ \Eprint
  {https://arxiv.org/abs/hep-th/0512261} {arXiv:hep-th/0512261} \BibitemShut
  {NoStop}%
\bibitem [{\citenamefont {Rose}\ \emph {et~al.}(2015)\citenamefont {Rose},
  \citenamefont {L\'eonard},\ and\ \citenamefont {Dupuis}}]{Rose:2015bma}%
  \BibitemOpen
  \bibfield  {author} {\bibinfo {author} {\bibfnamefont {F.}~\bibnamefont
  {Rose}}, \bibinfo {author} {\bibfnamefont {F.}~\bibnamefont {L\'eonard}},\
  and\ \bibinfo {author} {\bibfnamefont {N.}~\bibnamefont {Dupuis}},\ }\href
  {https://doi.org/10.1103/PhysRevB.91.224501} {\bibfield  {journal} {\bibinfo
  {journal} {Phys. Rev. B}\ }\textbf {\bibinfo {volume} {91}},\ \bibinfo
  {pages} {224501} (\bibinfo {year} {2015})},\ \Eprint
  {https://arxiv.org/abs/1503.08688} {arXiv:1503.08688 [cond-mat.quant-gas]}
  \BibitemShut {NoStop}%
\bibitem [{\citenamefont {Rose}\ \emph {et~al.}(2022)\citenamefont {Rose},
  \citenamefont {Pagani},\ and\ \citenamefont {Dupuis}}]{Rose:2021zdk}%
  \BibitemOpen
  \bibfield  {author} {\bibinfo {author} {\bibfnamefont {F.}~\bibnamefont
  {Rose}}, \bibinfo {author} {\bibfnamefont {C.}~\bibnamefont {Pagani}},\ and\
  \bibinfo {author} {\bibfnamefont {N.}~\bibnamefont {Dupuis}},\ }\href
  {https://doi.org/10.1103/PhysRevD.105.065020} {\bibfield  {journal} {\bibinfo
   {journal} {Phys. Rev. D}\ }\textbf {\bibinfo {volume} {105}},\ \bibinfo
  {pages} {065020} (\bibinfo {year} {2022})},\ \Eprint
  {https://arxiv.org/abs/2110.13174} {arXiv:2110.13174 [hep-th]} \BibitemShut
  {NoStop}%
\bibitem [{\citenamefont {Berges}\ \emph
  {et~al.}(2002{\natexlab{b}})\citenamefont {Berges}, \citenamefont
  {Tetradis},\ and\ \citenamefont {Wetterich}}]{Berges2002}%
  \BibitemOpen
  \bibfield  {author} {\bibinfo {author} {\bibfnamefont {J.~J.}\ \bibnamefont
  {Berges}}, \bibinfo {author} {\bibfnamefont {N.}~\bibnamefont {Tetradis}},\
  and\ \bibinfo {author} {\bibfnamefont {C.}~\bibnamefont {Wetterich}},\ }\href
  {https://doi.org/10.1016/S0370-1573(01)00098-9} {\bibfield  {journal}
  {\bibinfo  {journal} {Phys. Rept.}\ }\textbf {\bibinfo {volume} {363}},\
  \bibinfo {pages} {223} (\bibinfo {year} {2002}{\natexlab{b}})},\ \Eprint
  {https://arxiv.org/abs/hep-ph/0005122} {arXiv:hep-ph/0005122 [hep-ph]}
  \BibitemShut {NoStop}%
\bibitem [{\citenamefont {Canet}\ \emph
  {et~al.}(2003{\natexlab{a}})\citenamefont {Canet}, \citenamefont {Delamotte},
  \citenamefont {Mouhanna},\ and\ \citenamefont {Vidal}}]{Canet2003b}%
  \BibitemOpen
  \bibfield  {author} {\bibinfo {author} {\bibfnamefont {L.~L.}\ \bibnamefont
  {Canet}}, \bibinfo {author} {\bibfnamefont {B.}~\bibnamefont {Delamotte}},
  \bibinfo {author} {\bibfnamefont {D.}~\bibnamefont {Mouhanna}},\ and\
  \bibinfo {author} {\bibfnamefont {J.}~\bibnamefont {Vidal}},\ }\href
  {https://doi.org/10.1103/PhysRevB.68.064421} {\bibfield  {journal} {\bibinfo
  {journal} {Physical Review B - Condensed Matter and Materials Physics}\
  }\textbf {\bibinfo {volume} {68}},\ \bibinfo {pages} {64421} (\bibinfo {year}
  {2003}{\natexlab{a}})},\ \Eprint {https://arxiv.org/abs/hep-th/0302227}
  {arXiv:hep-th/0302227 [hep-th]} \BibitemShut {NoStop}%
\bibitem [{\citenamefont {Canet}\ \emph
  {et~al.}(2003{\natexlab{b}})\citenamefont {Canet}, \citenamefont {Delamotte},
  \citenamefont {Mouhanna},\ and\ \citenamefont {Vidal}}]{Canet2003}%
  \BibitemOpen
  \bibfield  {author} {\bibinfo {author} {\bibfnamefont {L.}~\bibnamefont
  {Canet}}, \bibinfo {author} {\bibfnamefont {B.}~\bibnamefont {Delamotte}},
  \bibinfo {author} {\bibfnamefont {D.}~\bibnamefont {Mouhanna}},\ and\
  \bibinfo {author} {\bibfnamefont {J.}~\bibnamefont {Vidal}},\ }\href
  {https://doi.org/10.1103/PhysRevD.67.065004} {\bibfield  {journal} {\bibinfo
  {journal} {Physical Review D - Particles, Fields, Gravitation and Cosmology}\
  }\textbf {\bibinfo {volume} {67}},\ \bibinfo {pages} {065004} (\bibinfo
  {year} {2003}{\natexlab{b}})}\BibitemShut {NoStop}%
\bibitem [{\citenamefont {Rosten}(2017)}]{Rosten:2014oja}%
  \BibitemOpen
  \bibfield  {author} {\bibinfo {author} {\bibfnamefont {O.~J.}\ \bibnamefont
  {Rosten}},\ }\href {https://doi.org/10.1140/epjc/s10052-017-5049-5}
  {\bibfield  {journal} {\bibinfo  {journal} {Eur. Phys. J.}\ }\textbf
  {\bibinfo {volume} {C77}},\ \bibinfo {pages} {477} (\bibinfo {year}
  {2017})},\ \Eprint {https://arxiv.org/abs/1411.2603} {arXiv:1411.2603
  [hep-th]} \BibitemShut {NoStop}%
\bibitem [{\citenamefont {Sonoda}(2015)}]{Sonoda:2015pva}%
  \BibitemOpen
  \bibfield  {author} {\bibinfo {author} {\bibfnamefont {H.}~\bibnamefont
  {Sonoda}},\ }\href {https://doi.org/10.1103/PhysRevD.92.065016} {\bibfield
  {journal} {\bibinfo  {journal} {Phys. Rev. D}\ }\textbf {\bibinfo {volume}
  {92}},\ \bibinfo {pages} {065016} (\bibinfo {year} {2015})},\ \Eprint
  {https://arxiv.org/abs/1504.02831} {arXiv:1504.02831 [hep-th]} \BibitemShut
  {NoStop}%
\bibitem [{\citenamefont {Rosten}(2019)}]{Rosten:2016zap}%
  \BibitemOpen
  \bibfield  {author} {\bibinfo {author} {\bibfnamefont {O.~J.}\ \bibnamefont
  {Rosten}},\ }\href {https://doi.org/10.1142/S0217751X19500271} {\bibfield
  {journal} {\bibinfo  {journal} {Int. J. Mod. Phys.}\ }\textbf {\bibinfo
  {volume} {A34}},\ \bibinfo {pages} {1950027} (\bibinfo {year} {2019})},\
  \Eprint {https://arxiv.org/abs/1605.01729} {arXiv:1605.01729 [hep-th]}
  \BibitemShut {NoStop}%
\bibitem [{\citenamefont {Polchinski}(1988)}]{polchinski1988scale}%
  \BibitemOpen
  \bibfield  {author} {\bibinfo {author} {\bibfnamefont {J.}~\bibnamefont
  {Polchinski}},\ }\href@noop {} {\bibfield  {journal} {\bibinfo  {journal}
  {Nuclear Physics B}\ }\textbf {\bibinfo {volume} {303}},\ \bibinfo {pages}
  {226} (\bibinfo {year} {1988})}\BibitemShut {NoStop}%
\bibitem [{\citenamefont {D'Attanasio}\ and\ \citenamefont
  {Morris}(1997)}]{DAttanasio:1997yph}%
  \BibitemOpen
  \bibfield  {author} {\bibinfo {author} {\bibfnamefont {M.}~\bibnamefont
  {D'Attanasio}}\ and\ \bibinfo {author} {\bibfnamefont {T.~R.}\ \bibnamefont
  {Morris}},\ }\href {https://doi.org/10.1016/S0370-2693(97)00866-6} {\bibfield
   {journal} {\bibinfo  {journal} {Phys. Lett.}\ }\textbf {\bibinfo {volume}
  {B409}},\ \bibinfo {pages} {363} (\bibinfo {year} {1997})},\ \Eprint
  {https://arxiv.org/abs/hep-th/9704094} {arXiv:hep-th/9704094 [hep-th]}
  \BibitemShut {NoStop}%
\bibitem [{\citenamefont {Balog}\ \emph
  {et~al.}(2019{\natexlab{b}})\citenamefont {Balog}, \citenamefont {Chat\'e},
  \citenamefont {Delamotte}, \citenamefont {Marohnic},\ and\ \citenamefont
  {Wschebor}}]{Balog:2019rrg}%
  \BibitemOpen
  \bibfield  {author} {\bibinfo {author} {\bibfnamefont {I.}~\bibnamefont
  {Balog}}, \bibinfo {author} {\bibfnamefont {H.}~\bibnamefont {Chat\'e}},
  \bibinfo {author} {\bibfnamefont {B.}~\bibnamefont {Delamotte}}, \bibinfo
  {author} {\bibfnamefont {M.}~\bibnamefont {Marohnic}},\ and\ \bibinfo
  {author} {\bibfnamefont {N.}~\bibnamefont {Wschebor}},\ }\href
  {https://doi.org/10.1103/PhysRevLett.123.240604} {\bibfield  {journal}
  {\bibinfo  {journal} {Phys. Rev. Lett.}\ }\textbf {\bibinfo {volume} {123}},\
  \bibinfo {pages} {240604} (\bibinfo {year} {2019}{\natexlab{b}})},\ \Eprint
  {https://arxiv.org/abs/1907.01829} {arXiv:1907.01829 [cond-mat.stat-mech]}
  \BibitemShut {NoStop}%
\bibitem [{\citenamefont {Simmons-Duffin}(2017)}]{Simmons-Duffin:2016wlq}%
  \BibitemOpen
  \bibfield  {author} {\bibinfo {author} {\bibfnamefont {D.}~\bibnamefont
  {Simmons-Duffin}},\ }\href {https://doi.org/10.1007/JHEP03(2017)086}
  {\bibfield  {journal} {\bibinfo  {journal} {JHEP}\ }\textbf {\bibinfo
  {volume} {03}},\ \bibinfo {pages} {086}},\ \Eprint
  {https://arxiv.org/abs/1612.08471} {arXiv:1612.08471 [hep-th]} \BibitemShut
  {NoStop}%
\bibitem [{\citenamefont {Campostrini}\ \emph {et~al.}(2002)\citenamefont
  {Campostrini}, \citenamefont {Hasenbusch}, \citenamefont {Pelissetto},
  \citenamefont {Rossi},\ and\ \citenamefont {Vicari}}]{Campostrini_2002}%
  \BibitemOpen
  \bibfield  {author} {\bibinfo {author} {\bibfnamefont {M.}~\bibnamefont
  {Campostrini}}, \bibinfo {author} {\bibfnamefont {M.}~\bibnamefont
  {Hasenbusch}}, \bibinfo {author} {\bibfnamefont {A.}~\bibnamefont
  {Pelissetto}}, \bibinfo {author} {\bibfnamefont {P.}~\bibnamefont {Rossi}},\
  and\ \bibinfo {author} {\bibfnamefont {E.}~\bibnamefont {Vicari}},\
  }\bibfield  {journal} {\bibinfo  {journal} {Physical Review B}\ }\textbf
  {\bibinfo {volume} {65}},\ \href {https://doi.org/10.1103/physrevb.65.144520}
  {10.1103/physrevb.65.144520} (\bibinfo {year} {2002})\BibitemShut {NoStop}%
\bibitem [{\citenamefont {Hasenbusch}(2022)}]{Hasenbusch_2022}%
  \BibitemOpen
  \bibfield  {author} {\bibinfo {author} {\bibfnamefont {M.}~\bibnamefont
  {Hasenbusch}},\ }\bibfield  {journal} {\bibinfo  {journal} {Physical Review
  B}\ }\textbf {\bibinfo {volume} {105}},\ \href
  {https://doi.org/10.1103/physrevb.105.054428} {10.1103/physrevb.105.054428}
  (\bibinfo {year} {2022})\BibitemShut {NoStop}%
\bibitem [{\citenamefont {Okabe}\ and\ \citenamefont {Oku}(1978)}]{Okabe78}%
  \BibitemOpen
  \bibfield  {author} {\bibinfo {author} {\bibfnamefont {Y.}~\bibnamefont
  {Okabe}}\ and\ \bibinfo {author} {\bibfnamefont {M.}~\bibnamefont {Oku}},\
  }\href {https://doi.org/10.1143/PTP.60.1287} {\bibfield  {journal} {\bibinfo
  {journal} {Progress of Theoretical Physics}\ }\textbf {\bibinfo {volume}
  {60}},\ \bibinfo {pages} {1287} (\bibinfo {year} {1978})},\ \Eprint
  {https://arxiv.org/abs/http://oup.prod.sis.lan/ptp/article-pdf/60/5/1287/5192266/60-5-1287.pdf}
  {http://oup.prod.sis.lan/ptp/article-pdf/60/5/1287/5192266/60-5-1287.pdf}
  \BibitemShut {NoStop}%
\bibitem [{\citenamefont {Vasil'ev}\ \emph {et~al.}(1982)\citenamefont
  {Vasil'ev}, \citenamefont {Pis'mak},\ and\ \citenamefont
  {Khonkonen}}]{Vasil'ev1982}%
  \BibitemOpen
  \bibfield  {author} {\bibinfo {author} {\bibfnamefont {A.~N.}\ \bibnamefont
  {Vasil'ev}}, \bibinfo {author} {\bibfnamefont {Y.~M.}\ \bibnamefont
  {Pis'mak}},\ and\ \bibinfo {author} {\bibfnamefont {Y.~R.}\ \bibnamefont
  {Khonkonen}},\ }\href {https://doi.org/10.1007/BF01015292} {\bibfield
  {journal} {\bibinfo  {journal} {Theoretical and Mathematical Physics}\
  }\textbf {\bibinfo {volume} {50}},\ \bibinfo {pages} {127} (\bibinfo {year}
  {1982})}\BibitemShut {NoStop}%
\bibitem [{\citenamefont {Broadhurst}\ \emph {et~al.}(1997)\citenamefont
  {Broadhurst}, \citenamefont {Gracey},\ and\ \citenamefont
  {Kreimer}}]{Broadhurst:1996ur}%
  \BibitemOpen
  \bibfield  {author} {\bibinfo {author} {\bibfnamefont {D.~J.}\ \bibnamefont
  {Broadhurst}}, \bibinfo {author} {\bibfnamefont {J.~A.}\ \bibnamefont
  {Gracey}},\ and\ \bibinfo {author} {\bibfnamefont {D.}~\bibnamefont
  {Kreimer}},\ }\href {https://doi.org/10.1007/s002880050500} {\bibfield
  {journal} {\bibinfo  {journal} {Z. Phys.}\ }\textbf {\bibinfo {volume}
  {C75}},\ \bibinfo {pages} {559} (\bibinfo {year} {1997})},\ \Eprint
  {https://arxiv.org/abs/hep-th/9607174} {arXiv:hep-th/9607174 [hep-th]}
  \BibitemShut {NoStop}%
\bibitem [{\citenamefont {Hu}(2001)}]{Hu_2001}%
  \BibitemOpen
  \bibfield  {author} {\bibinfo {author} {\bibfnamefont {X.}~\bibnamefont
  {Hu}},\ }\bibfield  {journal} {\bibinfo  {journal} {Physical Review Letters}\
  }\textbf {\bibinfo {volume} {87}},\ \href
  {https://doi.org/10.1103/physrevlett.87.057004}
  {10.1103/physrevlett.87.057004} (\bibinfo {year} {2001})\BibitemShut
  {NoStop}%
\bibitem [{\citenamefont {Clisby}\ and\ \citenamefont
  {D\"unweg}(2016)}]{Clisby16}%
  \BibitemOpen
  \bibfield  {author} {\bibinfo {author} {\bibfnamefont {N.}~\bibnamefont
  {Clisby}}\ and\ \bibinfo {author} {\bibfnamefont {B.}~\bibnamefont
  {D\"unweg}},\ }\href {https://doi.org/10.1103/PhysRevE.94.052102} {\bibfield
  {journal} {\bibinfo  {journal} {Phys. Rev. E}\ }\textbf {\bibinfo {volume}
  {94}},\ \bibinfo {pages} {052102} (\bibinfo {year} {2016})}\BibitemShut
  {NoStop}%
\bibitem [{\citenamefont {Clisby}(2017)}]{Clisby_2017}%
  \BibitemOpen
  \bibfield  {author} {\bibinfo {author} {\bibfnamefont {N.}~\bibnamefont
  {Clisby}},\ }\href {https://doi.org/10.1088/1751-8121/aa7231} {\bibfield
  {journal} {\bibinfo  {journal} {Journal of Physics A: Mathematical and
  Theoretical}\ }\textbf {\bibinfo {volume} {50}},\ \bibinfo {pages} {264003}
  (\bibinfo {year} {2017})}\BibitemShut {NoStop}%
\bibitem [{\citenamefont {Wiese}\ and\ \citenamefont
  {Fedorenko}(2019{\natexlab{a}})}]{Wiese:2019xmu}%
  \BibitemOpen
  \bibfield  {author} {\bibinfo {author} {\bibfnamefont {K.~J.}\ \bibnamefont
  {Wiese}}\ and\ \bibinfo {author} {\bibfnamefont {A.~A.}\ \bibnamefont
  {Fedorenko}},\ }\href {https://doi.org/10.1103/PhysRevLett.123.197601}
  {\bibfield  {journal} {\bibinfo  {journal} {Phys. Rev. Lett.}\ }\textbf
  {\bibinfo {volume} {123}},\ \bibinfo {pages} {197601} (\bibinfo {year}
  {2019}{\natexlab{a}})},\ \Eprint {https://arxiv.org/abs/1908.11721}
  {arXiv:1908.11721 [cond-mat.dis-nn]} \BibitemShut {NoStop}%
\bibitem [{\citenamefont {Wiese}\ and\ \citenamefont
  {Fedorenko}(2019{\natexlab{b}})}]{Wiese:2018dow}%
  \BibitemOpen
  \bibfield  {author} {\bibinfo {author} {\bibfnamefont {K.~J.}\ \bibnamefont
  {Wiese}}\ and\ \bibinfo {author} {\bibfnamefont {A.~A.}\ \bibnamefont
  {Fedorenko}},\ }\href {https://doi.org/10.1016/j.nuclphysb.2019.114696}
  {\bibfield  {journal} {\bibinfo  {journal} {Nucl. Phys.}\ }\textbf {\bibinfo
  {volume} {B946}},\ \bibinfo {pages} {114696} (\bibinfo {year}
  {2019}{\natexlab{b}})},\ \Eprint {https://arxiv.org/abs/1802.08830}
  {arXiv:1802.08830 [cond-mat.stat-mech]} \BibitemShut {NoStop}%
\end{thebibliography}%

\end{document}